\documentclass[fleqn,usenatbib]{mnras}

\usepackage[T1]{fontenc}
\usepackage{ae,aecompl}
\usepackage{graphicx}	
\usepackage{amsmath}	
\usepackage{amssymb}	
\usepackage{longtable}
\usepackage{multirow}
\usepackage{verbatim}
\usepackage{calc}
\usepackage{pdflscape}
\usepackage[flushleft]{threeparttable}
\usepackage[dvipsnames]{xcolor}

\newcommand{\ibh}{{\rm IBH}}
\newcommand{\bh}{{\rm BH}}
\newcommand{\bhs}{{\rm BHS}}
\newcommand{\gc}{{\rm GC}}
\newcommand{\gcz}{{\rm GC,0}}

\newcommand{\Ms}{~{\rm M}_\odot}
\newcommand{\Ls}{~{\rm L}_\odot}

\newcommand{\Log}{{\rm Log}}
\newcommand{\norm}[1]{\left\lVert#1\right\rVert}

\definecolor{orange}{RGB}{211, 112, 32}
\definecolor{cyan}{RGB}{29,214,222}
\definecolor{purple}{RGB}{181,12,141}
\definecolor{dgreen}{RGB}{56,181,12}

\title[IMBHs in Galactic globular clusters]{MOCCA-SURVEY Database I. Intermediate mass black holes in Milky Way globular clusters and their connection to supermassive black holes}
\author[M. Arca Sedda, A. Askar and M. Giersz]{Manuel Arca Sedda$^{1}$\thanks{E-mail:m.arcasedda@gmail.com}, Abbas Askar$^{2}$, Mirek Giersz$^{3}$
\\
$^{1}$Astronomisches Rechen-Institut, Zentrum f{\"u}r Astronomie, University of Heidelberg, M{\"o}nchhofstrasse 12-14, 69120, Heidelberg, Germany\\
$^{2}$Lund Observatory, Department of Astronomy, and Theoretical Physics, Lund University, Box 43, SE-221 00 Lund, Sweden\\
$^{3}$Nicolaus Copernicus Astronomical Center, Polish Academy of Sciences, 
ul. Bartycka 18, 00-716 Warsaw, Poland\\
}

\date{Accepted XXX. Received YYY; in original form ZZZ}
\pubyear{2017}

\begin{document}
\label{firstpage}
\pagerange{\pageref{firstpage}--\pageref{lastpage}}
\maketitle

\begin{abstract}
In this paper we explore the interplay between intermediate-mass black holes (IMBHs) and their nursing globular clusters (GCs), taking advantage of over 2000 Monte Carlo GC models. We find that the average density of IMBHs sphere of influence can be uniquely connected to the host GCs luminosity and half-light radius via a fundamental plane. We propose a statistical approach to systematically identify potential Galactic GCs harbouring either an IMBH or a massive subsystem comprised of stellar BHs. Our models show that the IMBH is often bound to a stellar companion or a stellar BH, which can lead to tidal disruption events or to low-frequency gravitational waves.
We show that GCs orbiting close to the Galactic Centre have a larger probability to witness IMBH formation during their early evolution. These low-orbit GCs can deliver several IMBHs into the galaxy innermost regions, with potential impact on both electromagnetic and GW emission. We discuss potential connections between IMBHs and SMBHs inhabiting galactic nuclei, exploring the possibility that in some cases they share similar formation pathways. 
\end{abstract}


\section{Introduction}
Intermediate mass black holes (IMBHs) constitute an elusive class of BHs with masses in the range $10^2-10^5\Ms$ that should fill the gap between stellar-mass and supermassive BHs \citep[see][for a recent review]{barack18}. Among others, a possible scenario for IMBH formation is through repeated stellar collisions in the innermost region of very dense stellar systems \citep{portegies00,zwart07,gaburov08,giersz15,mapelli16}. This requirement for high stellar densities makes globular clusters (GCs) promising sites for finding IMBHs. So far, a few Galactic GCs have been considered as potentially harbouring an IMBH \citep{silk75,bash08,maccarone08,lutz13,noyola10,lanzoni13,vandermarel10,strader2012,kamann14,mezcua17,kiziltan17,askar17b}, however, there is still no conclusive evidence for their presence in these clusters. 

The presence of an IMBH is often constrained via dynamical models tailored to reproduce the observed properties of a target GC. However, dynamical IMBH signatures can be efficiently mimicked by several effects, like rotation \citep{zocchi,zocchi17}, or the presence of a BH subsystem (BHS) in the cluster centre \citep{baumgardt03,peuten16,AS16,AAG18a,AAG18b,zocchi2019}.
In principle, kinematics of stars moving inside the putative IMBH sphere of influence, like radial velocity or proper motion, encode the information needed to describe the IMBH \citep{gebhardt05,noyola10,vandermarel10,kamann14}. However, following the motion of stars inside the IMBH's sphere of influence is challenging owing to their small size.
Another possible technique to measure IMBH properties relies upon millisecond pulsars (MSPs) timing
\citep{damico02,colpi02,colpi03,ferraro03, kiziltan17,perera2017,gieles18}. Although promising, also MSPs timing features do not provide conclusive proof for the existence of IMBHs. Two examples widely discussed in the literature are GCs 47 Tuc \citep{kiziltan17,abbate18,mann18}, and NGC 6624 \citep{perera2017,gieles18}.

Beside dynamical constraints, another viable possibility to ``see'' an IMBH inhabiting a GC centre is through accretion signatures. Potentially, an IMBH can accrete either i) stars' debris released during a tidal disruption event, ii)  pristine gas entrapped into the GC.

A tidal disruption event is triggered by stars passing too close to the IMBH, which are torn apart by tidal forces. Upon disruption, a fraction of stars' debris fall back onto the IMBH and feed it through an accretion disc, whose emission can be seen in different bands \- \citep{miller04b,shen14}. 

Subsequent X-ray emission is associated to the accretion disc orbiting the IMBH. Ultra-luminous X-ray sources (ULX) are thought to be the manifestation of such kind of events \citep{miller04}. First, because the kind of emission can be explained with stellar disruption operated by a moderately massive BHs, and second because they are usually observed either in the halo of early-type galaxies or close to, but clearly distinguishable from, the centre of late-type galaxies \citep{colbert02,miller04b}. Therefore, ULX are not necessarily associated with central supermassive black holes that inhabit galaxy nuclei. 

However, it must be noted that TDEs around IMBH are not the only suitable mechanisms to  explain ULX features. Indeed, in most observed cases the ULX is ascribed to emission from accreting neutron stars (NSs) \citep{wiktorowicz18}.

Although unique observational signatures associated with an IMBH have not yet been confirmed, several potential candidates have been observed in extragalactic globular clusters \citep{irwin06,shen19}. One example is the TDE observed in the young cluster MGG-11 in M82 \citep{kaaret01,matsumoto01}, which can be ascribed to an IMBH of mass $\sim 1000 \Ms$ \citep{hopman04,baumgardt06}.
More recently, \cite{Lin18} reported the discovery of a luminous X-ray outburst likely due to a TDE operated by an IMBH lurking in the centre of a massive star cluster at 12.5 kpc from the centre of the host, a lenticular galaxy. Such event might have originated by the disruption of a main sequence (MS) star \citep{chen18}.

A TDE triggered by an IMBH can involve not only MS stars, but also white dwarf (WDs) \citep{haas12,macleod14,fragione18c,anninos18,kawana18}. A WD-IMBH TDE can be characterised by peculiar features, like an underluminous thermonuclear explosion compared to standard SNIa, accompanied by a soft, transient X-ray signal \citep{rosswog08}. Also, WD disruption can lead to combined emission of X-rays and a burst of GWs \citep{haas12}. 

If the IMBH is accreting at a high rate,
multi-band radiation can be emitted by either a hot accretion disc or jets,  offering the possibility to constrain the IMBH mass via X-ray, optical and radio emission \citep{webb12,farrell14}. If the accretion rate is very low, the quiescent IMBH can appear as a radio continuum source thanks to synchrotron radio jets \citep{maccarone08,wrobel18}. However, so far observations of Galactic globular clusters have found no strong IMBH accretion signatures that could conclusively confirm their existence \citep{tremou18}. 

Along with electromagnetic emission, IMBHs are, in principle, also potential gravitational wave (GW) sources. A compact stellar remnant, like a neutron star (NS) or a stellar BH, passing sufficiently close to the IMBH can occasionally be captured on a tight binary, whose evolution is mostly driven by GW emission down to coalescence. These binaries are called intermediate-mass ratio inspirals, and are promising sources that can be observed with the next generation of spaced-borne gravitational wave detectors, like LISA \citep{seoane07,mandel08}. Ground-based observatories like LIGO and Virgo, instead, have the potential to observe mergers in the thin layer that separate stellar-mass and IMBHs, i. e. masses $O(100\Ms)$\citep{abbott17ibh}.

IMBHs can be connected to galactic nuclei, at least upon the assumption that they originally form in dense clusters. Indeed, star clusters are expected to slowly segregate toward galaxy centres due to dynamical friction, possibly contributing the formation of a nuclear cluster \citep{Trem75,Dolc93,AMB,antonini13,perets14,ASCD14a,ASCD14b}. Therefore, orbitally segregated globular clusters can transport their IMBHs into the galactic centre, possibly contributing to the SMBH development and growth \citep{ebisuzaki01,portegies06}. 
As recently investigated in a number of papers, delivered IMBHs can interact with each other and with the central SMBH, possibly leading to the formation of tight pairs that ultimately merge and release gravitational waves \citep{miller02,portegies06,ASCD17b, ASG16,mastrobuono14,fragione17}. 
Interestingly, a growing number of observations suggest that we already see IMBH candidates in the Milky Way centre \citep{oka17,takekawa17,ballone18,takekawa19}, whose origin might be related to infalling clusters. 

Due to the instrinsic difficulties in determining IMBHs existence, the development of reliable numerical technique capable of reproducing their formation is crucial to assess the actual probability for IMBHs to form in GCs. Recently, \cite{giersz15} took advantage of the MOCCA Monte Carlo code \citep{stdo86,giersz08,giersz13,hypki13} to create a database of over 2000 reliable GCs models. The MOCCA code features stellar evolution, taken into account via the binary stellar evolution synthesis tool \texttt{BSE} \citep{hurley2002}, and few body interactions via the \texttt{FEWBODY} integrator \citep{fregeau2004}.
The simulations sample, called MOCCA Survey Database I, covered a wide range of initial conditions in terms of GC initial mass and concentration, orbit, metallicity, binary fraction, SNe natal kicks \citep{askar17}. 
In nearly $20\%$ of the simulations, \cite{giersz15} found the formation of IMBHs with masses in the range $10^2-10^5\Ms$, triggered either by multiple stellar collisions in the earliest stage of GC evolution, or by slow accretion on a stellar BH seed that dominate the GC centre. In these regards, it must be noted that the current treatment for stellar collisions in MOCCA relies upon the assumption that $100\%$ of the star is accreted onto the BH\footnote{As it was shown in \citet{giersz15}, reduction of the amount of accreted mass onto an IMBH to only $25\%$ of the intruder mass do not have a significant effect on the IMBH mass grow rate.}. Improvements to such limitation are underway and will be presented in a forthcoming release. 
As discussed in our companion paper, $\sim 13\%$ of the models contain a long-lived BH subsystem (BHS) featuring up to a few hundreds of stellar mass BHs inhabiting the GC centre \citep{AAG18a}. Exploiting the database allowed us to demonstrate that it is possible to uniquely connect BH subsystems with GCs luminosity and observed core radius via a {\it fundamental plane} \citep{AAG18a}. 

Adapting our approach for the Milky Way GC system, we identified 29 possible targets that are, at present, harbouring $10-5\times 10^2$ BHs \citep{AAG18b,askar19}.  
As briefly discussed in our companion paper, a fundamental plane can also be defined for GCs harbouring an IMBH \citep{AAG18a}. In this paper, we make use of the MOCCA Survey Database I to deeply explore the connections between GCs and IMBHs from the theoretical and observational point of view.

On a small scale, we study the interplay between IMBHs and stars, investigating the possible development of TDEs and the formation of GW sources. On a GC scale, we investigate the possible formation of ``dark clusters'', namely GCs harbouring an IMBH that undergo an almost complete disruption, and the possible delivery of IMBHs in the Galactic Centre. On a global scale, we show that a handful scaling relations can be established between IMBH sphere of influence radius, density and mass, and the luminosity and core radius of the hosting GCs. Also, we develop a statistical technique to shortlist Galactic GCs harbouring either an IMBH or a BHS, which can be used to plan tailored observational campaigns.

The paper is organized as follows:  
in Section \ref{general} we present and discuss the general properties of our IMBH sample, in Section \ref{connection} we investigate the interplay between IMBHs and their hosts, and the potential connections with SMBHs in galactic nuclei, in Section \ref{scaling} we define a set of scaling relations connecting IMBHs with observational properties of the host GCs, in Section \ref{norm} we show how MOCCA models can be used to constrain the presence of an IMBH in Milky Way GCs, and in Section \ref{end} we summarize the main results of this work.

\section{General properties of IMBHs: influence radius and average density}
\label{general}

In galactic dynamics, the dynamical effect of a MBH sitting in the centre of a galaxy nucleus is quantified via the so-called influence radius, $R_\ibh$. This length scale encompasses the region of space where dynamics is substantially dominated by the MBH gravity. 
Indeed, $R_\ibh$ is defined as the radius at which the kinetic energy in the MBH surroundings equals the MBH potential well, i.e. $R_\ibh \sim GM_\ibh/\sigma^2$ \citep{peebles72}. If the MBH host phase space is well described by an isothermal sphere, this relation implies that the influence radius encompasses twice the MBH mass \citep{merritt04,merritt13}. From a dynamical point of view, an IMBH sitting in the centre of a massive globular cluster (GC) represents a somehow downsized version of a galactic nucleus, thus the influence radius associated to an IMBH can be defined as such that $M_\gc(R_\ibh) = 2M_\ibh$. Using the IMBH mass and the concept of sphere of influence, we define another important quantity, namely the IMBH {\it scale density}, $\rho_\ibh = 2M_\ibh/R_\ibh^3$, i.e. the average stellar density inside the influence sphere.

In this section, we show and discuss how these quantities are connected to each others. In the following, unless otherwise stated, the quantities taken into account are extracted from GCs models at 12 Gyr. All quantities referring to 12 Gyr have no pedix, while initial values are labelled with pedix $0$. 

\begin{figure}
    \centering
    \includegraphics[width=\columnwidth]{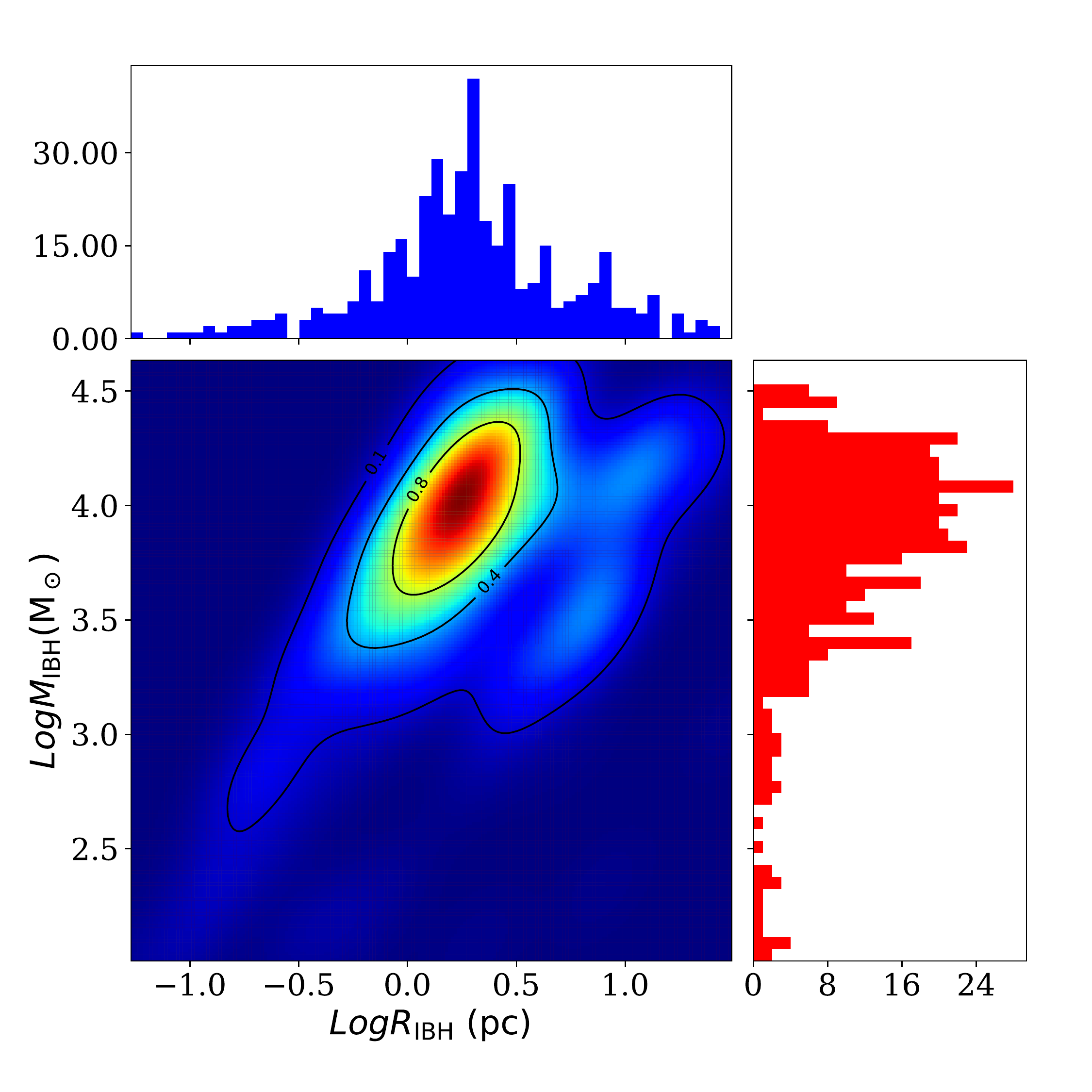}\\
    \includegraphics[width=\columnwidth]{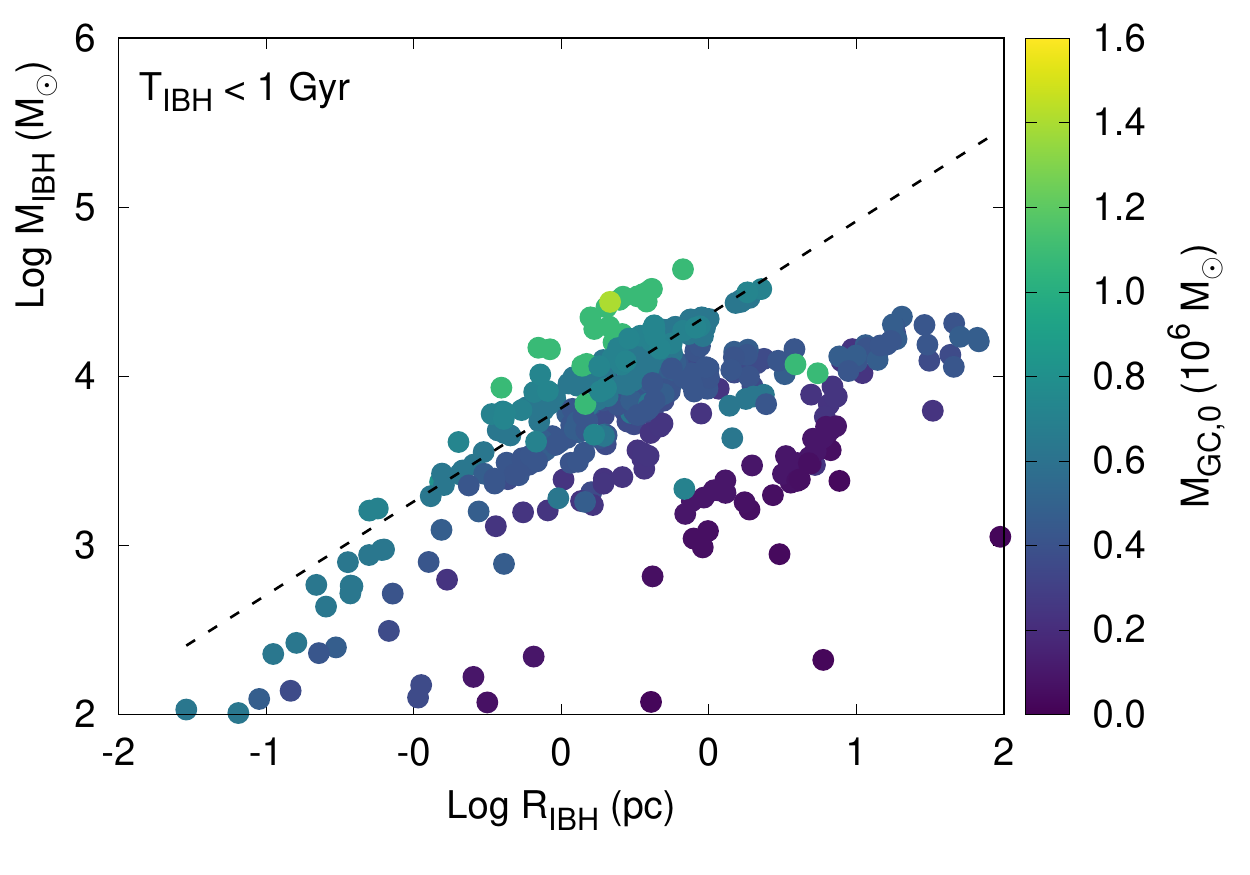}
    \caption{Top panel: IMBH mass as a function of its influence radius. Side histograms show the distribution of both quantities in our sample. Black contours identify where a fraction 0.1, 0.4, or 0.8 of all models lie. Bottom panel: same as in top panel, but we mark the host GC initial mass with different colours. The dotted line represent the best fit calculated taken into account only GCs with initial mass $M_\gcz \geq 5\times 10^5\Ms$.}
    \label{fig:mvsrinf}
\end{figure}

Figure \ref{fig:mvsrinf} shows the relation between IMBH mass and influence radius. It appears evident that the majority of IMBHs in MOCCA models have masses in the range $\Log (M_\ibh/\Ms) \simeq 3.5-4.5$ and  influence radii\footnote{for comparison, note that SgrA*, the Milky Way SMBH, has influence radius $R_{\rm Sgr~A*} \sim 1.5-3$ pc and mass $M_{\rm Sgr~A*} = 4\times 10^6\Ms$.} $R_\ibh = 1-3$ pc. 
Apart from this, we find two sub-populations of IMBHs particularly interesting. On one hand side, a handful of models show quite small IMBHs, with masses around $100-5000\Ms$ and compact spheres of influence, being $ R_\ibh \lesssim 0.32 {\rm ~pc}$. On the other hand side, a sizable fraction of models show large IMBH masses, $M_\ibh>10^3\Ms$, and influence radii exceeding $5-10$ pc. As we detail in the following, these sub-populations are characterised by a peculiar evolutionary history of the parent clusters. 

We find that the $M_\ibh-R_\ibh$ plane is inevitably connected with the host GC initial mass $M_\gc$, as shown in the bottom panel of Figure \ref{fig:mvsrinf}. Indeed, limiting the sample to GCs with initial masses above $M_\gcz \geq 5\times 10^5\Ms$, we find that the IMBH mass and the influence radius are connected via a tight powerlaw
\begin{equation}
    \Log M_\ibh = A\Log R_\ibh + B,
\end{equation}
being $A = 1.11\pm 0.05$ and $B = 3.81\pm 0.02$. Host GCs with a lower initial mass tend to deviate from this relation, moving toward larger influence radius. Interestingly, also the subsample of GCs having masses below $2\times 10^5\Ms$ shows a clear $M_\ibh-R_\ibh$ relation, with a slope $A' = 0.95\pm0.15$, thus compatible with the value that we find for larger GCs. 

\begin{figure}
    \centering
    \includegraphics[width=\columnwidth]{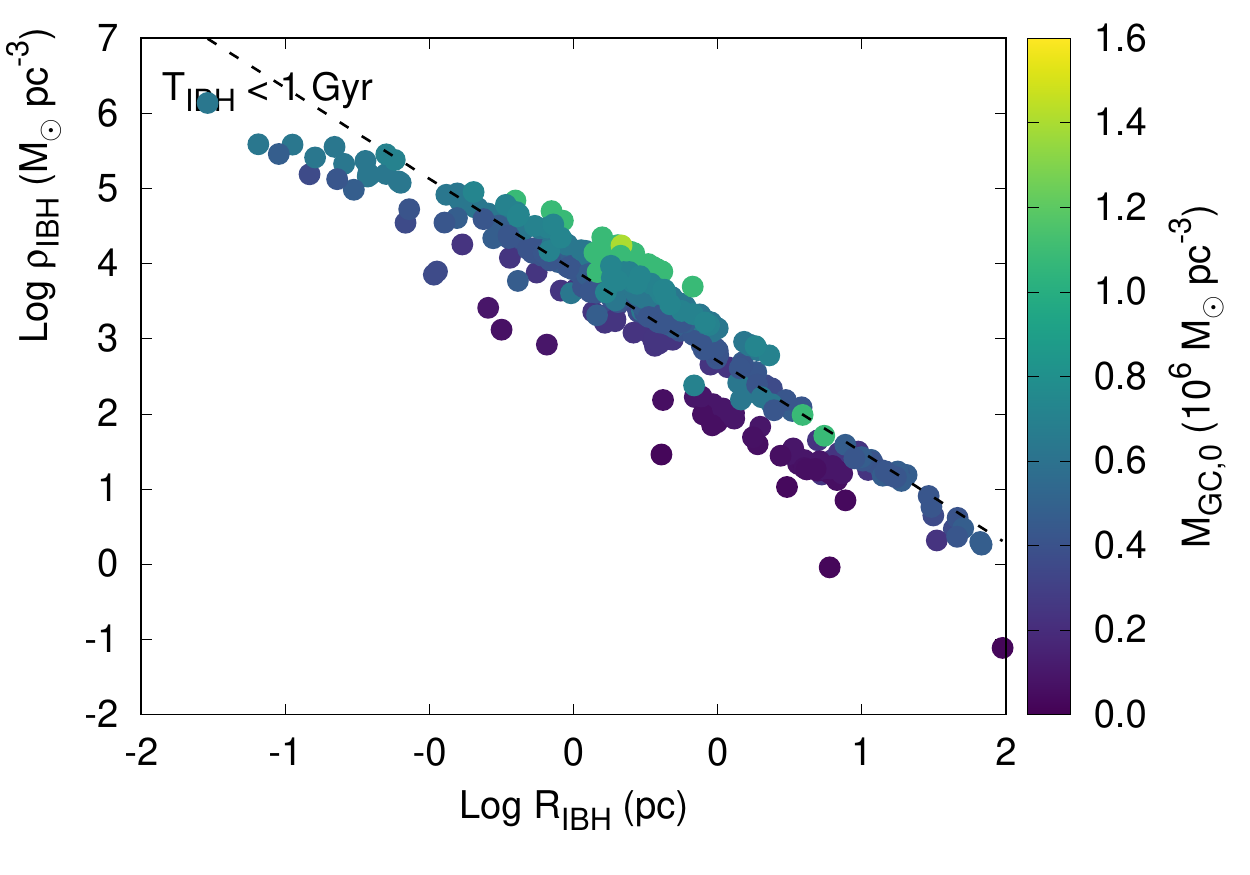}
    \caption{Stellar density $\rho_\ibh$ measured inside the influence radius, as a function of the influence radius. The color-coding marks the GC initial mass. Dotted black line marks the best fitting.}
    \label{fig:rhovsR}
\end{figure}

A further step needed to connect the IMBH with its surrounding is via a suitable relation between $R_\ibh$ and the average stellar density inside such radius, namely $\rho_\ibh$. As shown in Figure \ref{fig:rhovsR}, $\rho_\ibh$ is tightly related to the influence radius via a power-law
\begin{equation}
\Log \rho_\ibh = A\Log R_\ibh + B,
\end{equation}
with $A = -2.42\pm 0.05$ and $B = 3.92\pm 0.02$. As we show in the following section, $\rho_\ibh$ can be used to connect the GCs observational properties to the IMBH mass. 

The processes that drive IMBH formation in MOCCA can be divided into a ``FAST'' and a ``SLOW'' scenario \citep{giersz15}, which mostly differ in the timescale associated with the IMBH growth. In the FAST scenario the IMBH seed forms from direct collisions between stellar mass BHs and a very massive star built-up through main sequence stars collisions in the early stages of the star cluster lifetime. This usually happens in star clusters with initial central densities $\sim 10^8 \Ms$ pc$^{-3}$. As opposed to this, in the SLOW scenario the population of stellar-mass BHs is rapidly depleted via gravitational scattering until one or two BHs are left in the cluster grow. Over time-scales comparable to the GC core collapse time, the remaining BH starts growing via dynamical interactions resulting in binary mergers and mass transfer over $\gtrsim 10$ Gyr timescale.

\begin{figure}
    \centering
    \includegraphics[width=7cm]{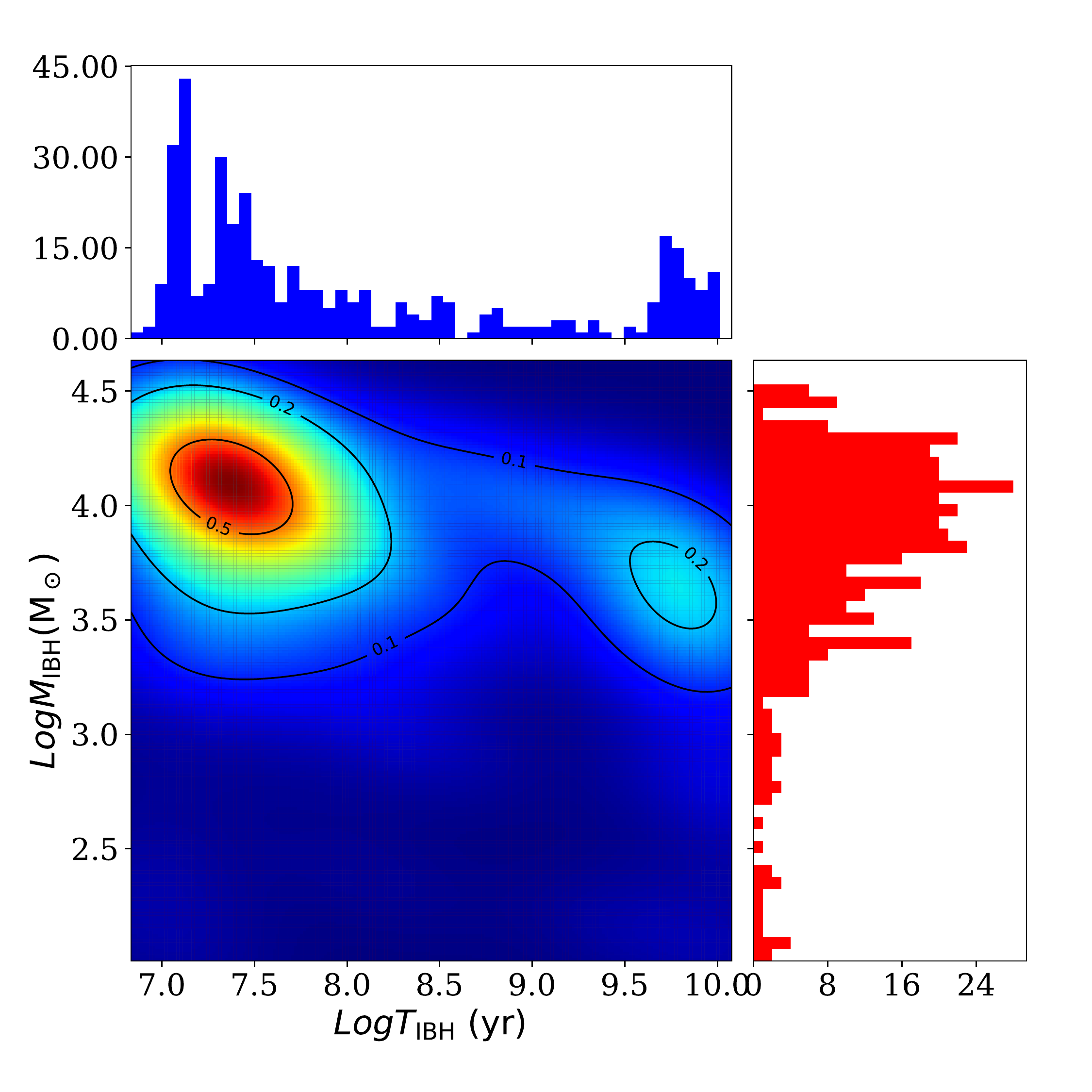}\\
    \includegraphics[width=7cm]{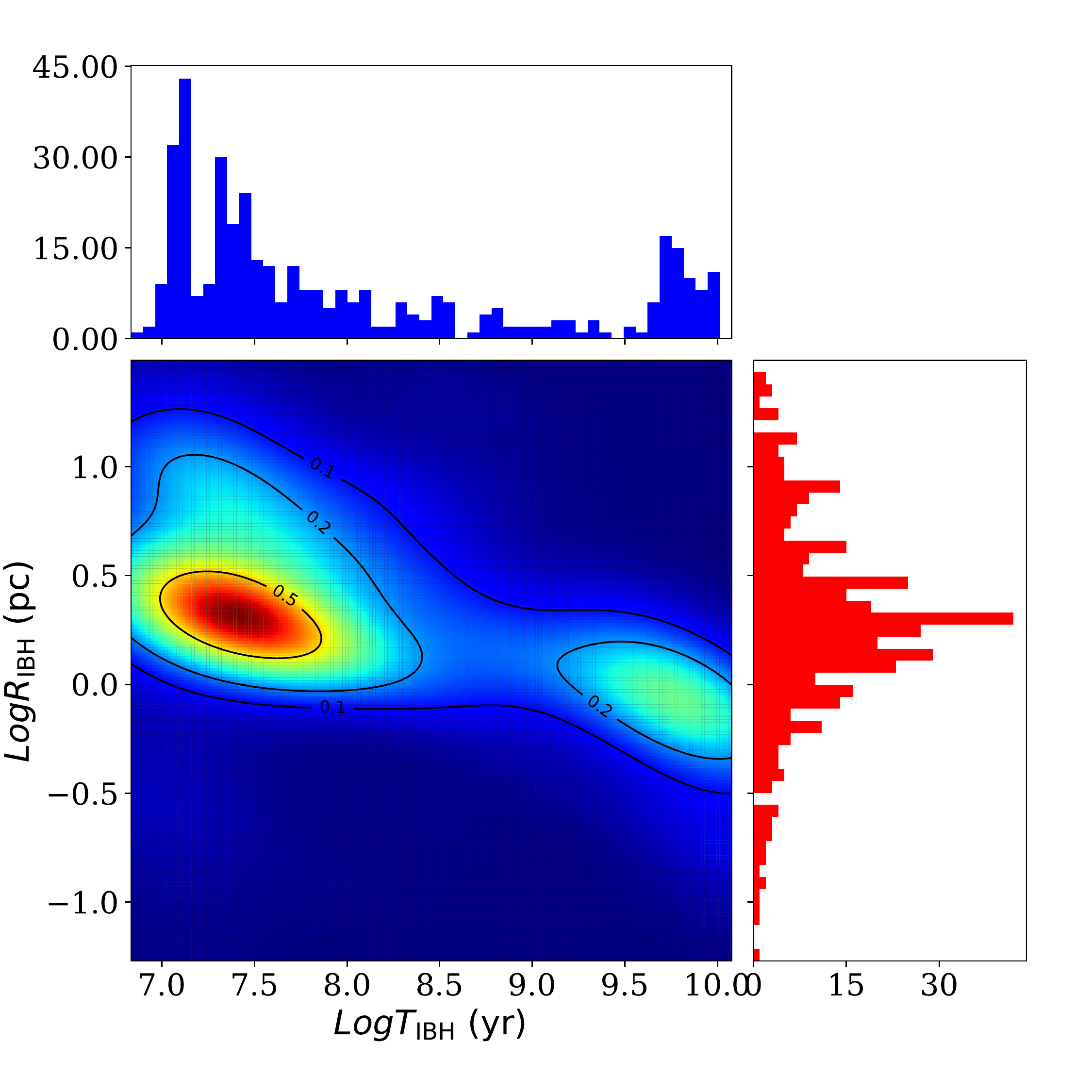}\\
    \includegraphics[width=7cm]{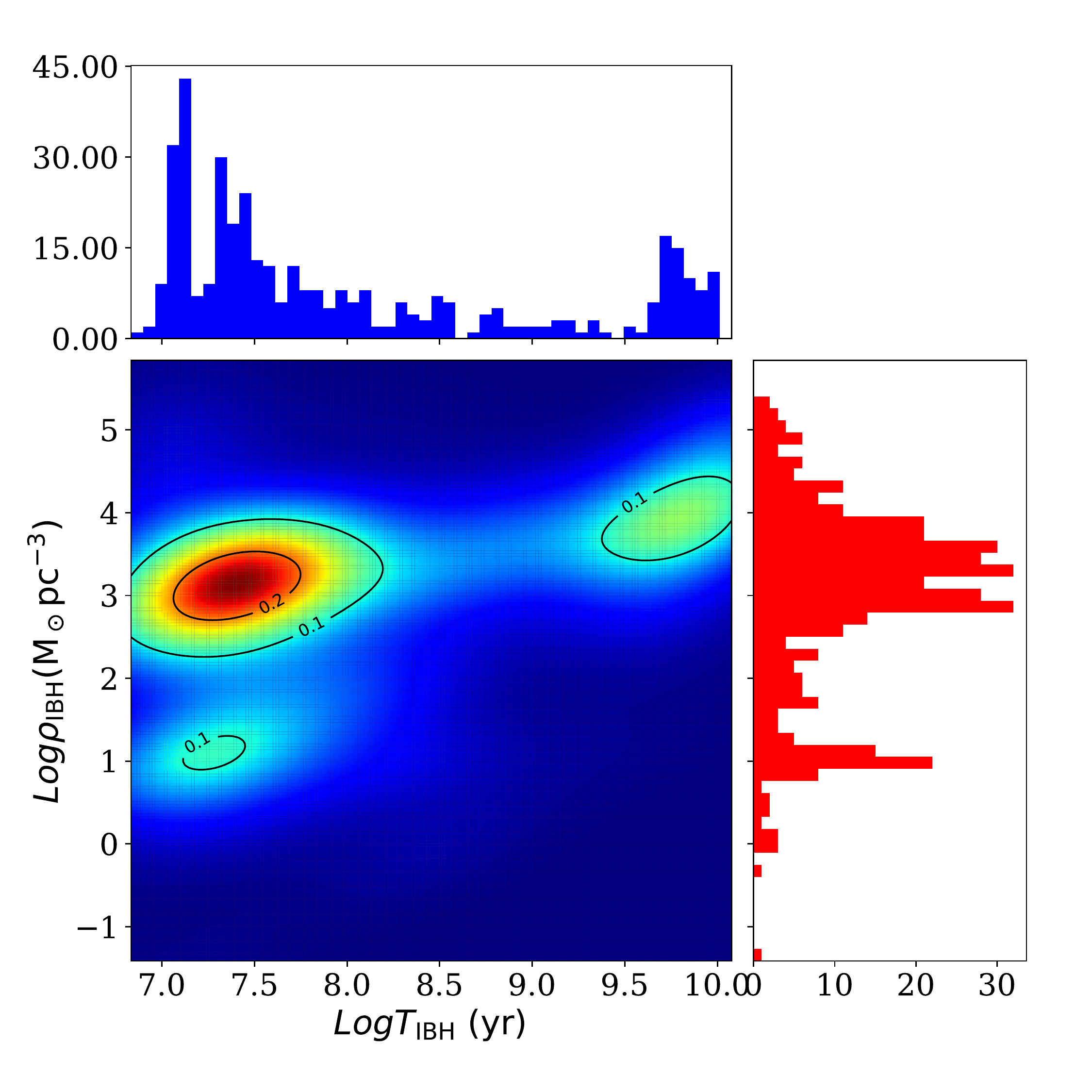}
    \caption{Time of IMBH formation as a function of the IMBH mass (top panel), influence radius (central panel) and average density (bottom panel).}
    \label{fig:formation}
\end{figure}

In order to quantify how fast the IMBH seed form, we define a {\it formation} time scale $T_\ibh$ as the time at which the IMBH mass exceeds a given threshold, namely $M_\ibh \geq 100\Ms$. Figure \ref{fig:formation} shows how $T_\ibh$ varies at varying $M_\ibh$ and $R_\ibh$. We note that IMBH mass and influence radius are taken at 12 Gyr, while $T_\ibh$ represents the first time the IMBH mass overtake $100 \Ms$.
The connection between the IMBH properties and its scale time is apparent. The $M_\ibh-T_\ibh$ plane clearly shows a bimodal trend in the top panel of the figure, with the majority of IMBHs forming on relatively short timescales $T_\ibh \simeq 10-30$ Myr. We note a weak anti-correlation, suggesting that IMBH forming via SLOW processes are lighter, on average. Also $R_\ibh$ shows a quite clear anti-correlation with $T_\ibh$ (central panel), being the sphere of influence smaller for IMBHs that form on longer timescales. Conversely, the stellar density within the influence radius increases at increasing the scale time, as seen in the bottom panel of Figure \ref{fig:formation}. This is due to the intrinsic differences between the FAST and SLOW mechanisms. Indeed, in the latter, the IMBH formation takes place after the host GC undergoes core collapse. As core collapse time scale is directly related to the GC density, IMBH forming through the SLOW scenario form in GCs that, at 12 Gyr, are characterized by relatively large densities. Conversely, in the FAST scenario the GC density is extremely high in the first stages, and trigger the rapid formation and growth of the IMBH. As the IMBH grows quickly, it shapes strongly the host GC, leading in many cases to an apparent expansion of the core, leading the GCs to have, at 12 Gyr, quite sparse matter distribution and a significant mass loss.
Therefore, the three panels can be interpreted, at least in part, as a direct consequence of the dynamical effect that the IMBH has on its surroundings. 
An IMBH that forms on a short timescale, of the order of a few tens Myr, and reaches a mass as high as $O(1\%)$ the host GC mass 
represents a very efficient energy source. The net energy flux injected to passing by stars is balanced by the host GC core expansion, which triggers a decrease in the local density and reduces the interaction rate.

\begin{figure}
    \centering
    \includegraphics[width=\columnwidth]{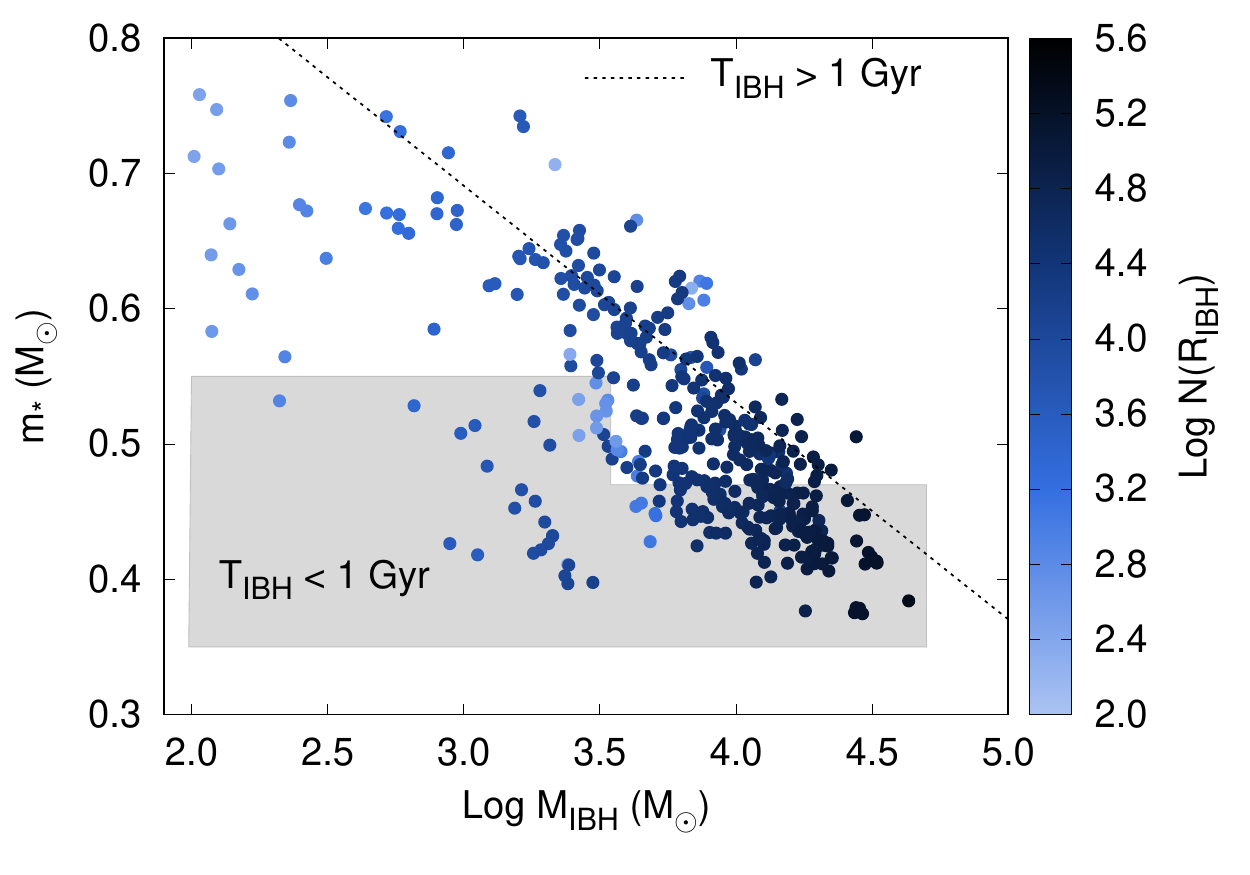}
    \caption{Average mass of stars inside the influence radius as a function of the IMBH mass. The number of stars enclosed within the IMBH influence radius is color-coded. 
    The lower part of the shaded region contains only IMBH with $T_\ibh < 1$ Gyr, while in the upper part both FAST and SLOW IMBH are distributed. Dotted black line identifies the scaling relation connecting $m_*$ and the subsample of ``SLOW IMBHs'' ($T_\ibh>1$ Gyr) with mass above $M_\ibh \geq10^3\Ms$.}
    \label{fig:maver}
\end{figure}

The presence of an IMBH in the GC centre can also affect the properties of the stellar population inhabiting its vicinity. In order to explore such effect, we calculate the average stellar mass inside the IMBH influence radius as $2M_\ibh/N(R_\ibh)$, being $N(R_\ibh)$ the number of stars enclosed within $R_\ibh$. This quantity is shown in Figure \ref{fig:maver}, as a function of the IMBH mass and the number of stars orbiting inside the IMBH sphere of influence. We note an evident anticorrelation between $m_*$ and the IMBH mass. Heavier IMBHs are associated to lower $m_*$ values, while at increasing the number of stars orbiting the IMBH the offset of such relation shifts to higher values, meaning that smaller GCs have, on average, smaller stars in the IMBH surroundings. Moreover, we find that there is a well defined region of the $M_\ibh-m_*$ plane populated only by IMBHs with short growth time, namely ($T_\ibh < 1$ Gyr). In the remaining part of the plane, both FAST and SLOW IMBHs coexist. If we limit our analysis to SLOW IMBHs only and assuming $M_\ibh>1000\Ms$, we found that the average stellar mass and the IMBHs mass are related through a powerlaw of form
\begin{equation}
    \Log \left(\frac{M_\ibh}{\Ms}\right) = -0.16\left(\frac{m_*}{\Ms}\right) + 1.18.
\end{equation}
These results suggest that IMBHs forming quickly are, on average, associated to a lighter stellar population compared to SLOW IMBHs.

This is likely due to the fact that FAST IMBHs ($T_\ibh < 1$ Gyr), affect dynamics over more than $10$ Gyr. Interactions with the IMBH will involve the heaviest stars, which sink to the centre via mass segregation and then are kicked out via dynamical interactions. This can easily lower the average stellar mass. Conversely, SLOW IMBHs are formed after the collapse time, thus the cluster core is expected to be still dense and containing the most massive stars due to mass segregation. However, in this case IMBHs have more moderate masses, from $\sim 10^2\Ms$ to a few $10^3\Ms$.

\section{IMBHs and their hosts}
\label{connection}
The interplay between global GCs evolution and the development of an IMBH in their centre is still partly obscure. In this section, we try to explore the relations that link GCs evolutionary paths to the main properties of their IMBHs. 

\subsection{ Formation of binary systems containing an IMBH }

Out of the 407 MOCCA models containing an IMBH, $\sim 20\%$ of the entire MOCCA Survey database I, 
we serendipitously discover several models in which the IMBH is part of a binary at 12 Gyr. 

Binary systems comprised of an IMBH and a close stellar companion represent one of the most promising kind of sources that can be used to directly observe the IMBH. Indeed small and repeated perturbations caused by surrounding stars can drive the companion toward an orbit that closely approaches the IMBH. Depending on its stellar type, the companion can either undergo disruption, with consequent electromagnetic emission, or be captured on an orbit that slowly spiral in due to the emission of gravitational waves until coalescence.

In the first case, the IMBH is outshined by a strong flare in the X-rays \citep{shen14,miller04b}, followed by emission in a wide portion of the electromagnetic spectrum. 
Therefore, providing a comprehensive set of information about IMBHs residing in a binary has paramount importance to place constraints on a variety of astrophysical phenomena. Another important effect, not explored in this paper, arises from possible direct collisions between stars and the IMBH, which can be copious especially in the first phases of IMBH growth.

\begin{figure}
    \includegraphics[width=\columnwidth]{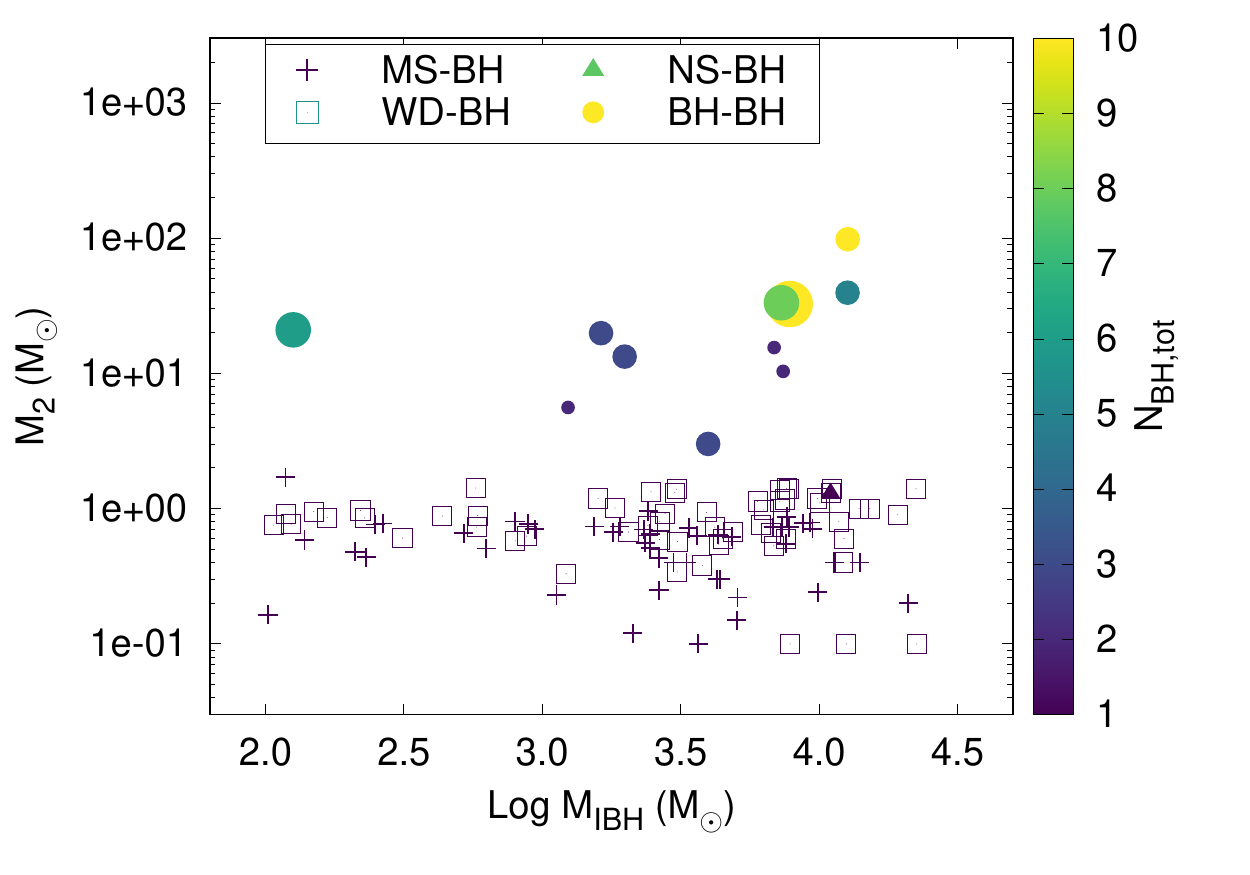}
    \caption{Upper limit to the companion mass as a function of the IMBH mass. Color-coding identifies the total number of BHs harboured in the hosting GCs at 12 Gyr. We differentiate between MS-BH (cross), WD-BH (open square), NS-BH (triangles), and BH-BH (circles) pairs. For BHs, larger circles correspond to a larger total number of BHs (single + binary), with the largest corresponding to 9 BHs and the smallest corresponding to 2.}
    \label{fig:IBHinBin}
\end{figure}

In our sample, at 12 Gyr we find 56 MS-IMBH, 66 WD-IMBH, 1 NS-IMBH and 11 BH-IMBH binaries. Figure \ref{fig:IBHinBin} shows the mass of the components for different companion stellar types. Assuming that our models are representative of actual GC, we find that an IMBH developing in its central regions has a probability to form a binary of $32.8\%$, either with a MS star ($13.7\%$), a WD ($16.2\%$), a NS ($0.2\%$), or a BH ($2.7\%$). We stress here that the probability is computed taking as global sample the 407 MOCCA models that contains at 12 Gyr an IMBH with mass larger than $100\Ms$.

Interestingly, we discover IMBH paired with a MS, WD, or NS only in clusters containing no stellar BHs at 12 Gyr. This is expected by the fact that a number of stellar mass BHs orbiting around the IMBH would efficiently scatter low-mass stars, pumping energy in the IMBH surroundings and preventing its pairing with smaller stars. 

This occurs in the early evolutionary phases, as BHs are the first object segregating into the cluster core due to their mass. When IMBH kicked out most of the BHs via multiple scatterings, lower mass stars migrate to the centre and start interacting with the IMBH as well. 
The simultaneous presence of an IMBH and a few BHs is a signature of the FAST process, as in the SLOW scenario usually all but one or two BHs are ejected before the IMBH grows up.
Therefore, the absence of stellar BHs in presence of an IMBH-stellar pair has crucial observational implications. Indeed, observing a TDE associated to star disruption from an IMBH would immediately provide us with a further clue, that the host cluster with high probability does not contain stellar BHs. The orbit of GCs containing an IMBH-stellar binary are distributed in a wide range of galactocentric distances, being $R_\gc = 1-30$ kpc. This suggests that TDEs triggered by an IMBH can be seen in both the inner and outskirt of the host galaxy. Such picture is clearly compatible with both observations of ULX in the vicinity of galaxy centres and with off-centered X-ray flares, like the one recently observed by \citep{Lin18}.

On the other hand, models containing a BH-IMBH binary are initially located at Galactocentric distances peaked around 10 kpc, except for one model for which $R_\gc = 3$ kpc. This kind of binary are particularly interesting as they are potential GW emitters and might be observed by the next generation of GW detectors like LISA \citep{seoane07} and, at some extent, with current ground-based observatories, like LIGO \citep{abbott17ibh}. In Table \ref{tab:catalogue}, we provide a catalogue of our BH-IMBH binaries, showing the binary masses, the total number of BHs, and the number of single and binary BHs. 
 At 12 Gyr, we find that the number of BHs is relatively small, $N_\bh = 2-9$, due to the fact that most of them are merged or kicked out during the IMBH assembly phases. The scarce BH population is characterised by a very high binary fraction, though, being the ratio between the number of binary and single BHs at 12 Gyr $N_{\rm BHB}/N_\bh = (0.5-1)$ in all the cases. We find that BH-IMBH binaries are characterized by a broad total mass distribution in the range $M_\bh+M_\ibh = (1.5\times 10^2-1.3\times 10^4)\Ms$ and mass ratios $M_\bh / M_\ibh = (7.6\times 10^{-4}-0.17)$.
 In all cases but one, the IMBH form via the FAST scenario over typical timescales $T_\ibh = 10-160$ Myr. In one case, the IMBH forms at a later stage, being $T_\ibh\simeq 10$ Gyr. This is somehow expected from the peculiar properties of the two formation channels discussed here, as in the SLOW scenario the IMBH growth starts when the BH reservoir is practically emptied, thus making more difficult for the IMBH to pair with a stellar BH.

\begin{table}
    \centering
    \caption{Main parameters of BH-IMBH binaries in MOCCA: name of the source, IMBH and BH mass, total number of BHs in the host GC at 12 Gyr, number of single and binary BHs, IMBH formation time}
   \begin{tabular}{cccccccc}
    \hline
    \hline
        NAME & $M_\ibh$ & $m_2$ & $ N_\bh $ &$ N_{\rm bin} $&$ N_{\rm sin} $ & $T_\ibh$ \\
             &  ($\Ms$) &($\Ms$)&           &               &      &    (Myr)  \\
    \hline
        GW01 & 1240.3&  5.6& 2& 2& 0& $10^4$\\
        GW02 & 7417.5& 10.4& 2& 2& 0& $12.4$\\
        GW03 & 6881.0& 15.5& 2& 2& 0& $157.1$\\
        GW04 & 1630.8& 19.8& 3& 2& 1& $12.8$\\
        GW05 & 1984.6& 13.3& 3& 2& 1& $10.7$\\
        GW06 & 3972.4&  3.0& 3& 2& 1& $10.3 $\\
        GW07 &12699.0& 98.1& 4& 2& 2& $21.9$\\
        GW08 &12684.0& 39.5& 4& 2& 2& $21.6$\\
        GW09 &  126.0& 21.0& 5& 4& 1& $9.9$\\
        GW10 & 7312.0& 33.3& 7& 4& 3& $35.4$\\
        GW11 & 7835.0& 32.6& 9& 6& 3& $23.2$\\
    \hline
    \end{tabular}
    \label{tab:catalogue}
\end{table}

\subsection{On the IMBH dynamical influence on the host evolution:  Globular versus Dark Clusters}

\begin{figure}
    \centering
    \includegraphics[width=8cm]{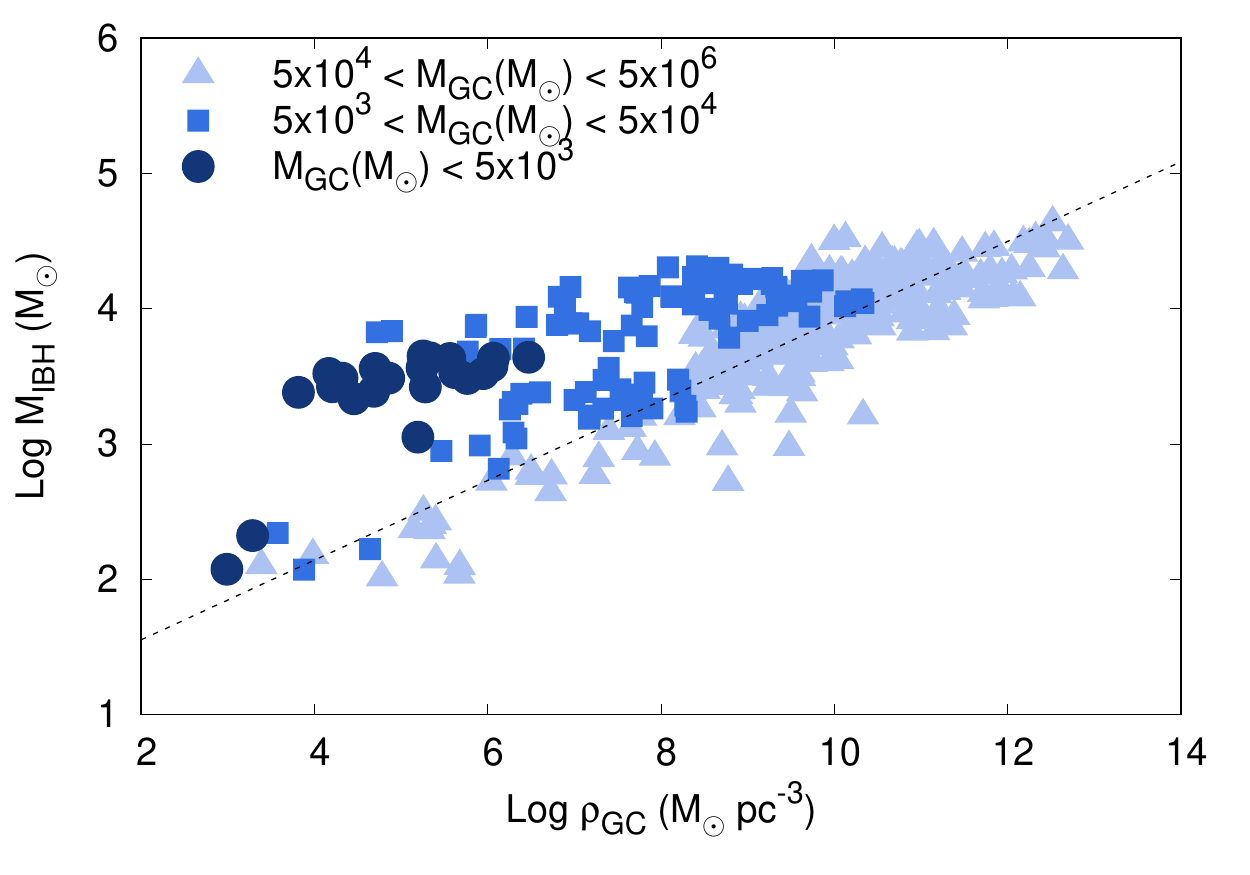}\\
    \includegraphics[width=8cm]{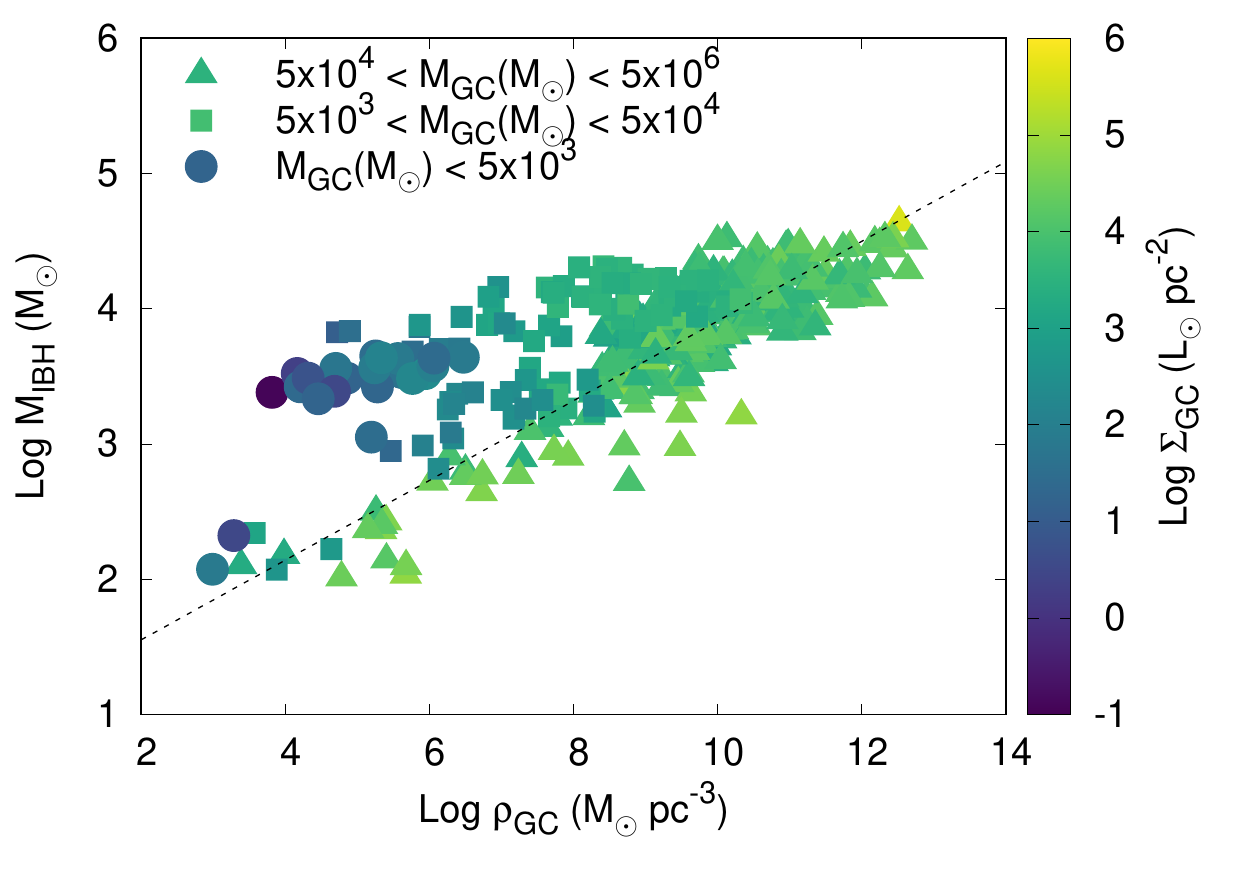}\\
    \includegraphics[width=8cm]{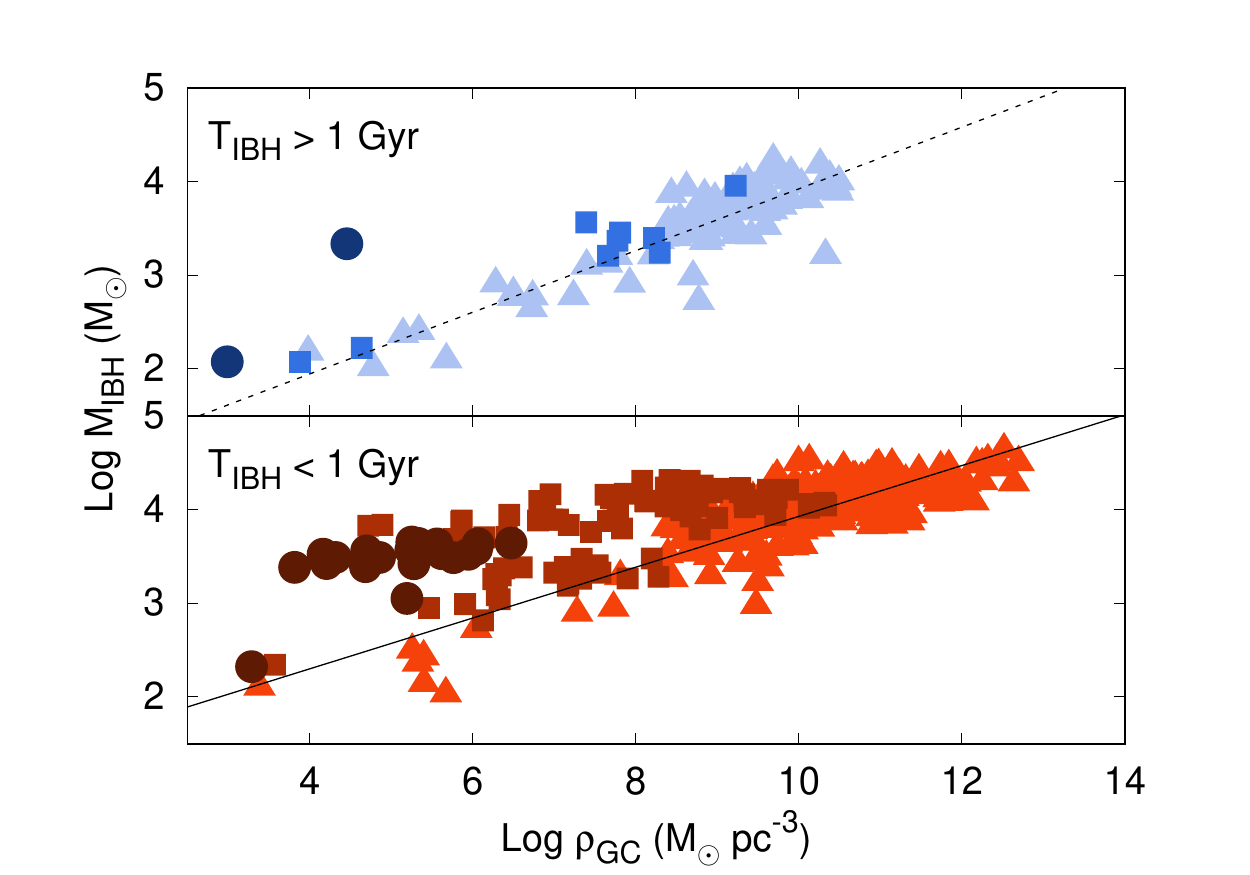}
    \caption{Top panel: IMBH mass as a function of the GC present-day central density. Different colors label different GC mass ranges: $M_{\gc}<10^3\Ms$ (dark blue), $10^3<M_{\gc}/\Ms<10^4$ (blue), $M_{\gc}>10^4\Ms$ (light blue). Central panel: same as above, but in this case we use color-coding to identify GCs surface brightness calculated at 12 Gyr. Bottom panel: same as above, but dividing the population into SLOW $T_\ibh>1$ Gyr and FAST IMBHs $T_\ibh<1$ Gyr. In all panels, the dotted black line marks the best fit of the subsample of models having $M_{\rm GC}>10^4\Ms$.}
    \label{fig:gcrelibh}
\end{figure}

In order to obtain a straightforward set of relations that can be used to place constraints on the presence of an IMBH in observed GCs, we need to explore the host properties at present time\footnote{We assume $t=12$ Gyr as present time, for coeherence's sake in comparison with previous works that used the same set of simulations.}. Figure \ref{fig:gcrelibh} shows how the IMBH mass correlates with the host cluster central density at 12 Gyr. In the figure we divide the GCs sample in three population, according to different ranges of present-day mass values: $M_{\gc}<10^3\Ms$, $10^3<M_{\gc}/\Ms<10^4$, $M_{\gc}>10^4\Ms$. If we restrict the analysis to the population of clusters that preserve a mass compatible with GCs typical masses, i.e. $M_{\gc}>10^4\Ms$, we find a tight correlation that allows to convert the GC central density to the IMBH mass, being 
\begin{equation}
\Log M_\ibh = \alpha\Log \rho_{\gc } + \beta,
\end{equation}
with $\alpha = 0.295 \pm0.009$ and $\beta = 0.96 \pm0.09$. 

Clusters that deviate from this relation are particularly interesting, as in these systems internal processes, stellar evolution, and the IMBH dynamical influence caused a very efficient mass removal. As a consequence, GCs outside the sequence are characterised by an IMBH-to-GC mass ratio very high, being $M_\ibh/M_{\gc}>0.1-1$. We can define these ``dark clusters'', as their stellar mass is comparable to the mass in dark remnants \citep[see also][]{askar17b}. As outlined in the central panel, dark clusters are characterised by lower values of the central surface brightness, compared to normal GCs. Different behaviours are also associated to different IMBH formation channels. As shown in the bottom panel of Figure \ref{fig:gcrelibh}, GCs that undergo FAST IMBH growth and with masses at 12 Gyr below $M_\gc < 5\times 10^4\Ms$ significantly deviate from the relation. Such deviation seems much less significant for GCs hosting ``SLOW'' IMBHs, although in this case the number of small final GC masses is evidently smaller. 
This is due again to the fact that a SLOW IMBH form right after GC core collapse, thus implying that the IMBH is only a small fraction of the host total mass, even if the cluster core collapse time is comparable to the cluster dissolution time.

\subsection{On the delivery of IMBHs harboured in low-orbit GCs}

\begin{figure}
    \centering
    \includegraphics[width=\columnwidth]{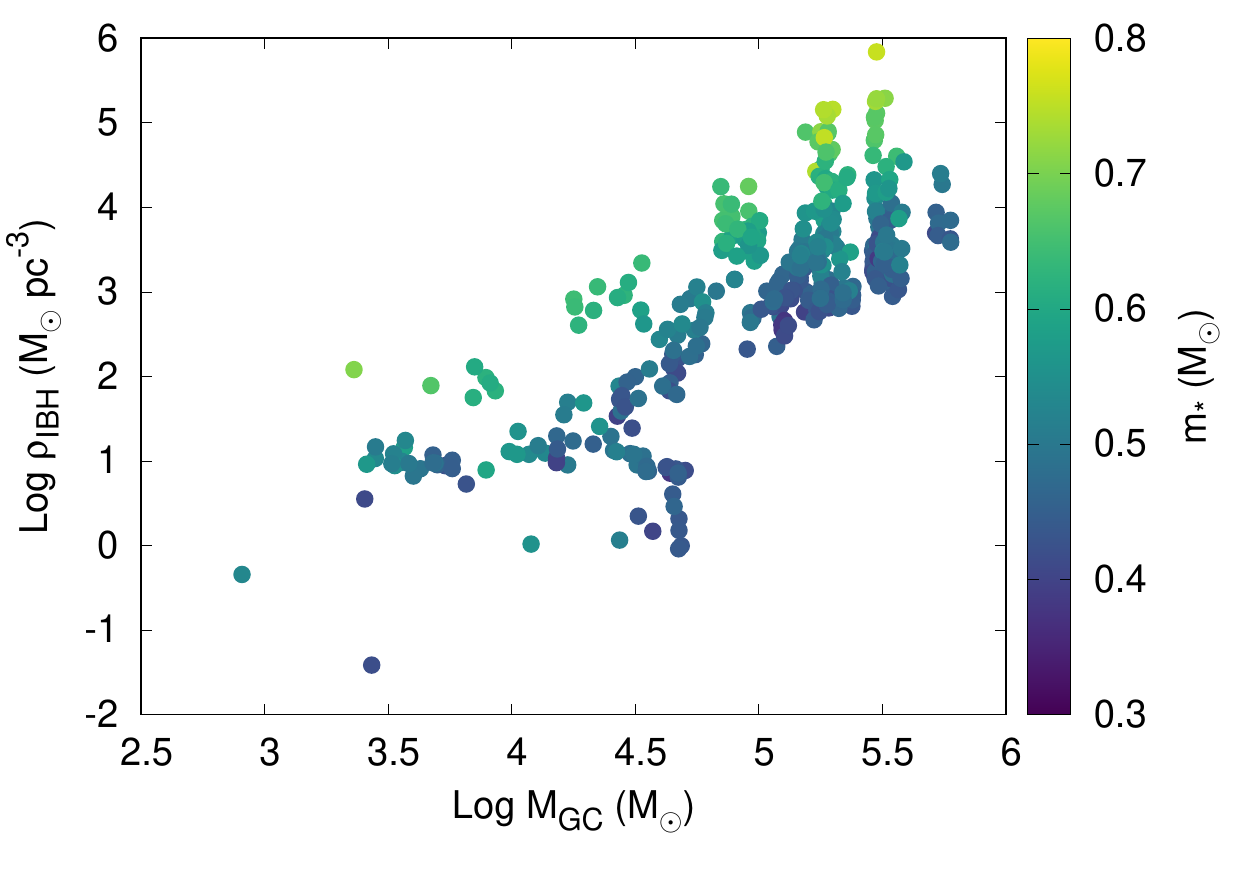}
    \caption{Stellar density inside $R_\bh$ as a function of the GC mass calculated at 12 Gyr. Coloured maps identify different values of the average stellar mass.}
    \label{fig:gc12}
\end{figure}

The stellar density in the IMBH surroundings, measured through the $\rho_\ibh$ parameter, seems to correlate with the host cluster mass calculated at 12 Gyr, as shown in Figure \ref{fig:gc12}. Although loosely tight, this relation shows that heavier GCs are expected to harbour IMBHs with denser spheres of influence. Also, at fixed the GC mass, the larger the $\rho_\ibh$ the heavier the average stellar mass measured within $R_\ibh$ and the smaller the sphere of influence. 

\begin{figure}
    \centering
    \includegraphics[width=\columnwidth]{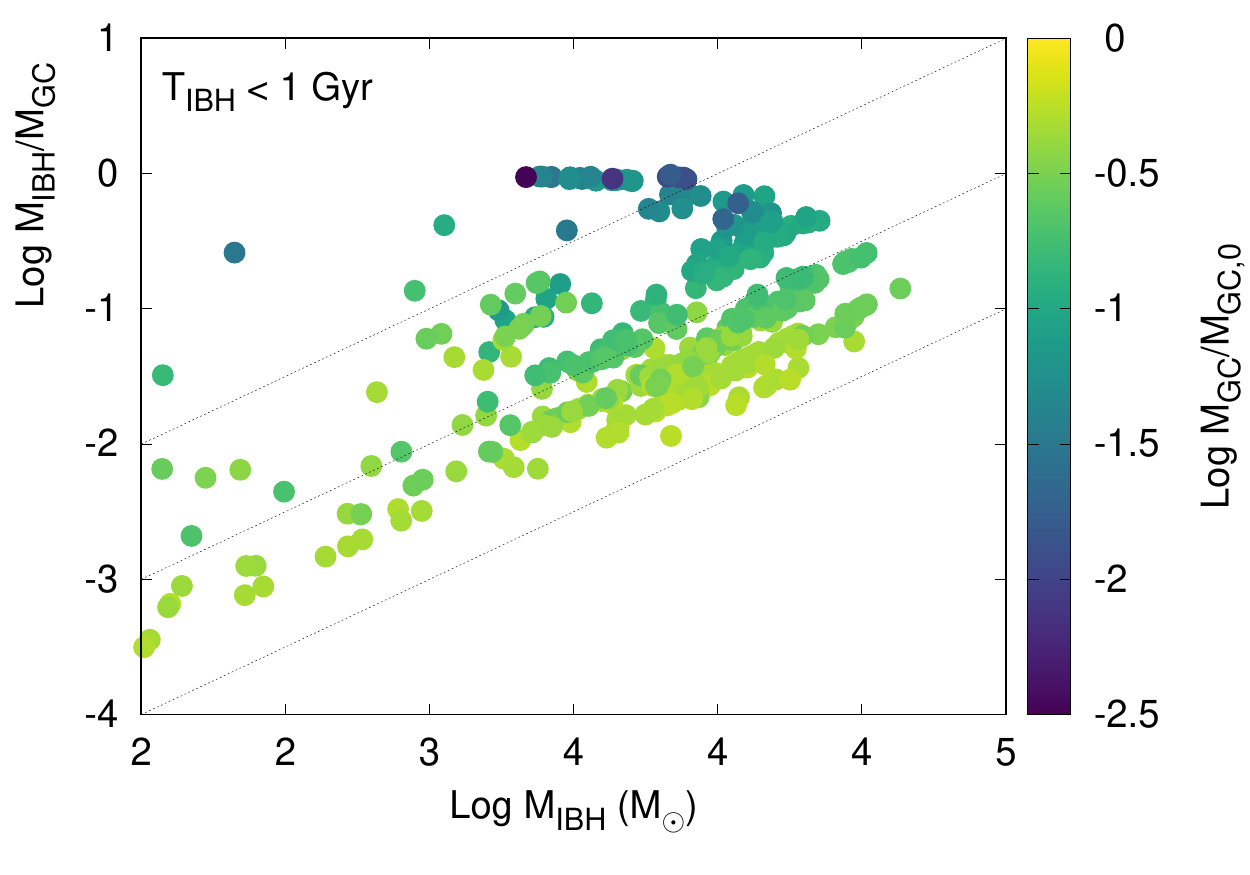}
    \caption{Ratio between the IMBH and the GC masses as a function of the IMBH mass. The colour-coding labels the final to initial GC mass ratio. Dotted lines mark the linear relation $M_\ibh/M_{\gc}\propto M_\ibh$. All the quantities are calculated at 12 Gyr.}
    \label{fig:ibhovergc}
\end{figure}

Figure \ref{fig:ibhovergc} shows one of the basic properties, namely the IMBH-to-GC mass ratio measured at 12 Gyr as a function of the IMBH mass and the ratio between the final and initial value of the GC mass. In the case in which mass loss processes have a marginal impact on the GC evolution, $M_{\gc}\gtrsim 0.3M_\gc$, the IMBH-GC mass ratio lies in the range $M_\ibh/M_{\gc} \simeq 10^{-4}-10^{-1.5}$. However, GCs experiencing a more effective mass loss, $M_{\gc}\lesssim 0.1 M_\gc$, have two main properties: i) they produces heavy IMBHs, with masses $ > (0.5-3)\times10^4\Ms$, and ii) the IMBH mass is $90-100\%$ the host GC mass. 
These are the dark clusters discussed in the previous section, i.e. a type of clusters containing more or less $10^4$ stars, whose mass budget is completely dominated by the IMBH. 

\begin{figure}
    \centering
    \includegraphics[width=\columnwidth]{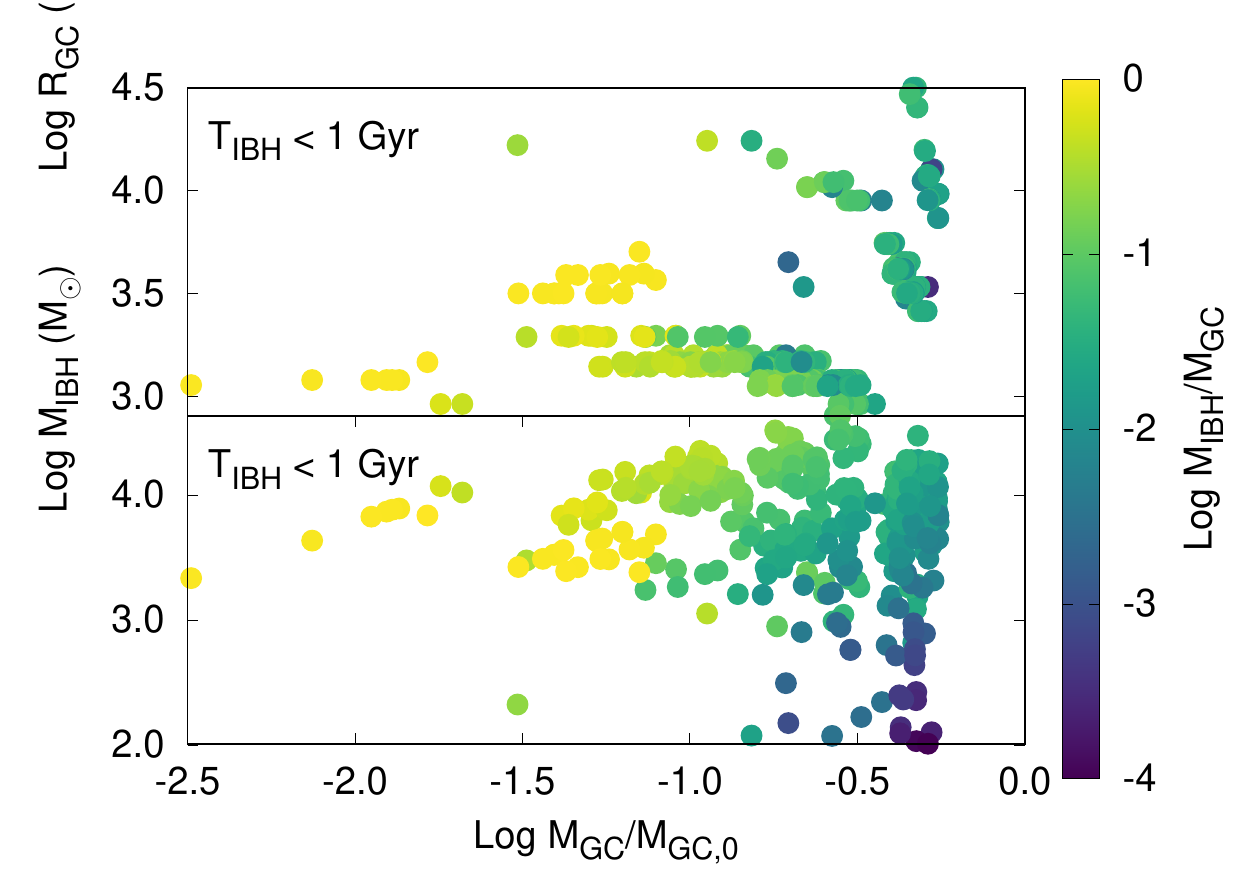}
    \caption{Galactocentric distance (top panel) and IMBH mass (bottom panel) as a function of GC mass measured at 12 Gyr, normalized to its initial value. Different colours mark the IMBH-to-GC mass ratio measured at 12 Gyr. }
    \label{fig:ibhmult}
\end{figure}

In Figure \ref{fig:ibhmult}, we show the correlation between the GC Galactocentric distance and the GC final mass, normalized to its initial value. First, we note that most GCs suffering an efficient mass removal, i. e. $M_{\gc} \leq 0.1M_\gc$, move on orbits within 1-10 kpc from the Galaxy Centre, as seen in the top panel. Second, disrupted GCs bring with them quite heavy IMBHs, with masses $M_\ibh\simeq 10^3-10^4\Ms$. 

The two points above can have interesting consequences for the evolution of galaxy nuclei. Indeed, as discussed in \cite[see their Figure 1]{AAG18a}, several MOCCA models containing an IMBH have dynamical friction times smaller than a Hubble time, suggesting that these IMBHs can be delivered into the Galactic Centre. 
Following \cite{ASCD15He} \citep[but see also][]{ASCD14a}, we calculate the dynamical friction (df) timescale for all the MOCCA models having an IMBH. We find 189 systems with a df time smaller than 12 Gyr. Among them, in 157 cases the IMBH growth time is much shorter than the df time, thus implying that the IMBH already grew-up when the parent cluster arrives in the Galactic Centre. Therefore, our analysis suggest that some GCs orbiting in the inner kpc of the host galaxy can witness the birth and growth of an IMBH and subsequently deliver it into the central galactic regions via df-driven inspiral. Some of these IMBHs can be wandering at present days in the MW centre, as some observations suggest \citep{oka17,takekawa17,takekawa19}, or might have merged in the past few Gyr with the Galactic SMBH \citep{ASG16}. On the other hand, to many IMBHs freely orbiting the inner 10 pc would leave dynamical imprints on the nuclear cluster that are not compatible with observations \citep{mastrobuono14}, thus implying that the potential number of IMBHs moving in there is most likely $< 10$.

\subsection{What is the connection between intermediate-mass and supermassive black holes?}

As shown in the previous section, our MOCCA models show that stellar collisions, either occurring in the GC early life, or on secular time-scales, can buildup an IMBH with masses as high as a few $10^4\Ms$. 
This process allows the formation of IMBHs that perfectly fit the gap between stellar mass and supermassive black holes (SMBHs). 

\begin{figure*}
    \centering
    \includegraphics[width=15cm]{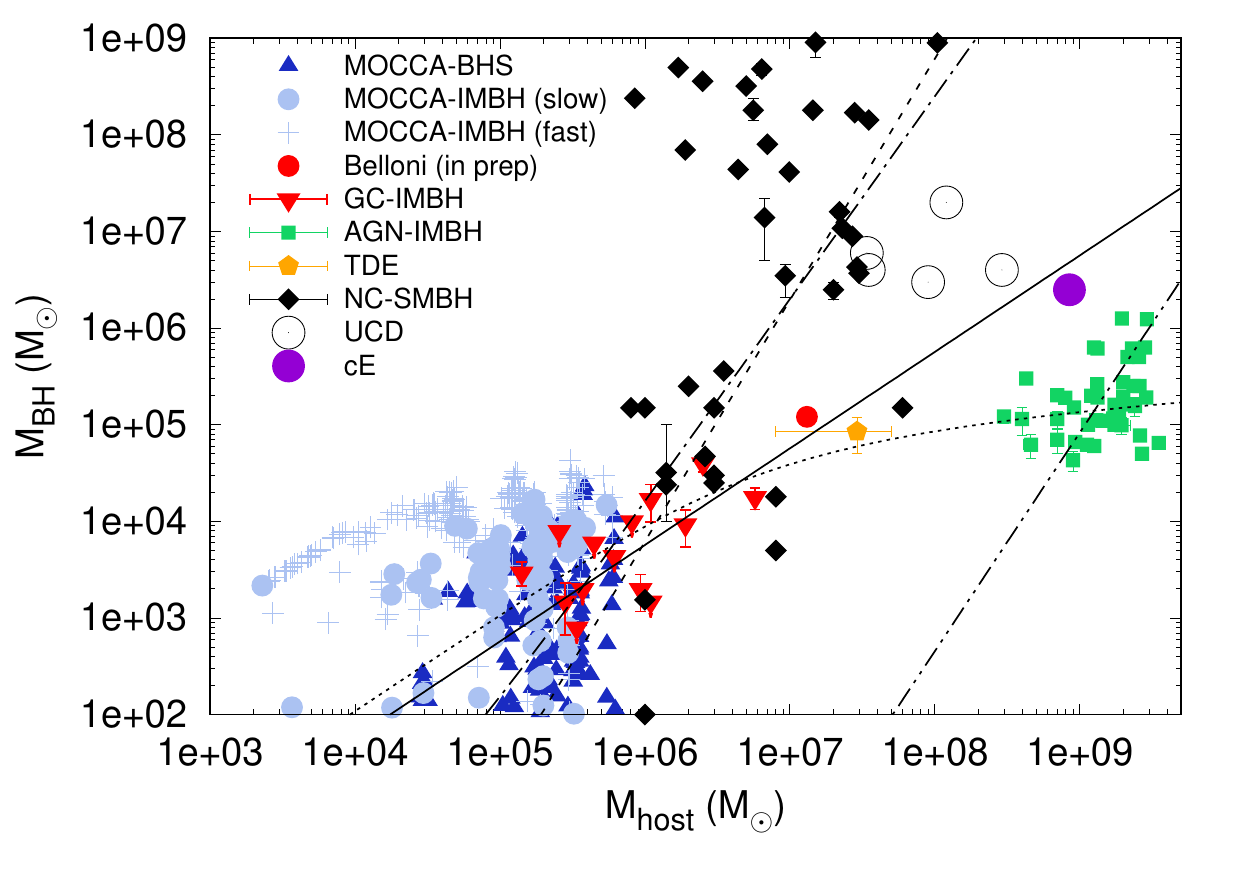}
    \caption{Central dark object mass as a function of the host mass for MOCCA models containing an IMBH (light blue filled points), or a BH subsystem (open blue points). We also add a new simulation performed with MOCCA and tailored to represent a NSC (red filled circle, Diogo Belloni, private communication).
    For comparison, we show the sample of Galactic GCs possibly harbouring an IMBH compiled by \citet{Lutzgendorf13} (red triangles), a sample of galactic nuclei containing a supermassive BH \citep{graham09,Neum12,reines16, Kormendy13} (black diamond), a sample of low-luminosity AGN containing bona fide IMBHs \citep{chilingarian18} (green squares), and the star cluster hosting an IMBH candidate that originated the X-ray flare observed by \citet{Lin18}. 
    We overlay the GC-BHS relation proposed by \citet{AS16} (straight black line), the BH-NC relation proposed by \citet{graham16b} (dashed line), the BH-bulge mass relation for low-mass galaxies \citep{graham15} (double dotted-dashed line) 
    and two possible fits that link either GC to galactic nuclei (dash-dotted line) or to low-luminosity AGNs (dotted line).}
    \label{fig:mscaling}
\end{figure*}

As for IMBHs, also the formation of supermassive black holes is a partly unsolved mystery of modern astronomy. At moment, the most credited scenario for SMBH growth are either via stellar collisions between pop III stars, or via monolithic collapse of a gaseous cloud \citep[see][for a review]{barack18}. The first scenario -- mergers among pop III stars -- is quite similar to what occurs in GCs harbouring an IMBH. Therefore, it might be interesting to explore whether it is possible to establish a connection between IMBHs and SMBHs. In order to compare MOCCA models with observations, we build a compilation of data available in literature for different hosts. In particular we consider:
\begin{itemize}
    \item the sample of 13 IMBH candidates found in Galactic GCs provided by \citep{Lutzgendorf13};
    \item the sample of SMBHs and NCs that are known to co-exist in several galaxies, like the MW \citep{graham09,Neum,Kormendy13};
    \item the IMBH and host cluster mass estimates provided by \cite{Lin18}, which are based on a X-ray flare originate during the tidal disruption of a passing by star;
    \item the compilation of bona fide IMBHs observed in a population of low-luminosity AGN \citep{reines16,chilingarian18};
    \item a few data available for ultra-compact dwarfs (UCDs) and for the M32 compact elliptical galaxy (cE).
\end{itemize}
The aforementioned sample is shown in Figure \ref{fig:mscaling}.
such a sample contains very different objects, likely characterized by different formation and evolution histories. 

Comparing MOCCA data with the few observational constraints available for MW globulars show a relatively good agreement between models and observations, although it must be noted that the present-day GC mass in MOCCA SURVEY DATABASE I never exceeds $10^6\Ms$, making difficult the comparison with heavier observed GCs. Also, the population of MOCCA models containing an BHS, as defined in \cite{AAG18a}, seem to complement IMBH-dominated systems, arranging in a similar region of the $M_\bh-M_{\rm host}$ plane.

We find two possible relations that fits relatively well MOCCA IMBHs, one leading to a connection with SMBHs in NCs
\begin{equation}
    \Log M_\ibh = M_{\rm i1}\left(\frac{M_{\rm cl}}{M_{\rm c1}} + 1 \right)^{\gamma}
    \label{smbhnc}
\end{equation}
with $M_{\rm c0}=2\times 10^6\Ms$ and $M_{\rm i0}=5\times 10^4\Ms$ (dash-dotted line in Figure \ref{fig:mscaling}), and the other with SMBH in low-mass AGNs:
\begin{equation}
    \Log M_\ibh = M_{\rm i0}\Log \left(\frac{M_{\rm cl}}{M_{\rm c0}} +1\right), 
    \label{smbhagn}
\end{equation}
with $M_{\rm i1} = 1\Ms$, $M_{\rm c1} = 10^4\Ms$, and $\gamma = 2.2$ (dotted line in Figure \ref{fig:mscaling}).  Note that in both equations we used $M_{\rm cl}$ to indicate indifferently GCs and NCs. Both relations provide a good match to observed putative IMBHs. 

Regarding the IMBH-SMBH-NC connection -- Equation \ref{smbhnc} -- we note that the relation provides quite a good fit also to SMBH masses in the $10^4-10^7\Ms$ region. Also, host cluster mass loss may play a role, as this mechanism is expected to be more effective in GCs than in galactic nuclei. Heavier SMBHs tend to deviate from the correlation. This might be sign of different concurrent formation processes. For low-mass SMBH, star collisions or stellar feeding onto a massive seed could have dominated the SMBH growth process, while heavier SMBH might have growth mostly because of gaseous accretion.

The IMBH-AGN connection -- Equation \ref{smbhagn} -- instead, provides a good matching with galaxies hosting low-mass AGNs. We note that this relation matches the recently observed TDE event detected by \cite{Lin18} in an extragalactic star cluster. Therefore, the relation suggests that a possible formation channel for low-mass AGN in dwarf galaxies is dominated by stellar collisions and accretion onto a heavy seed. It must be noted that such formation scenario seems to be supported also from the observational point of view, being the host of low-mass AGN all characterized by compact and bright nuclei \citep{chilingarian18}. 

Unfortunately, MOCCA GCs in the current database have present-day masses below $\sim 10^6\Ms$, thus limiting the phase space region available for our analysis. Future investigations will allow us to reach larger GC masses, possibly helping in shedding light on the bridge between IMBH-SMBH.
Heavier MOCCA models are currently running and under analysis. In one of them, tailored to model a NSC sitting in the centre of its host galaxy, represented with a mass $M_{\rm NSC} = 1.31\times 10^7\Ms$ and half-mass radius $r_h = 1$ pc, we find the formation of a MBH with mass $M_\ibh = 1.21\times 10^5\Ms$ (Diogo Belloni, private communication). Such point places excellently in between the scaling relation provided by \cite{AS16}, connecting both BHS and IMBHs with their host masses, and the scaling relation connecting IMBHs and low-mass AGN depicted here. 

\section{Observational scaling relations connecting IMBHs and their hosts}
\label{scaling}
Unveiling a unique way to connect an IMBH with some observational properties of the hosting cluster represents one of the most challenging quests in modern astrophysics.

For all MOCCA models containing an IMBH, we calculated the mass to light ratio, defined as the ratio between GCs present day mass and total visual luminosity, namely $M_{\gc}/L_{V}$. We find that in 25 models, i.e. $\simeq 6\%$ the sample, this quantity exceeds 10. These extremely low luminosity systems are the dark clusters discussed in the previous section.

\begin{figure*}
    \centering
    \includegraphics[width=\textwidth]{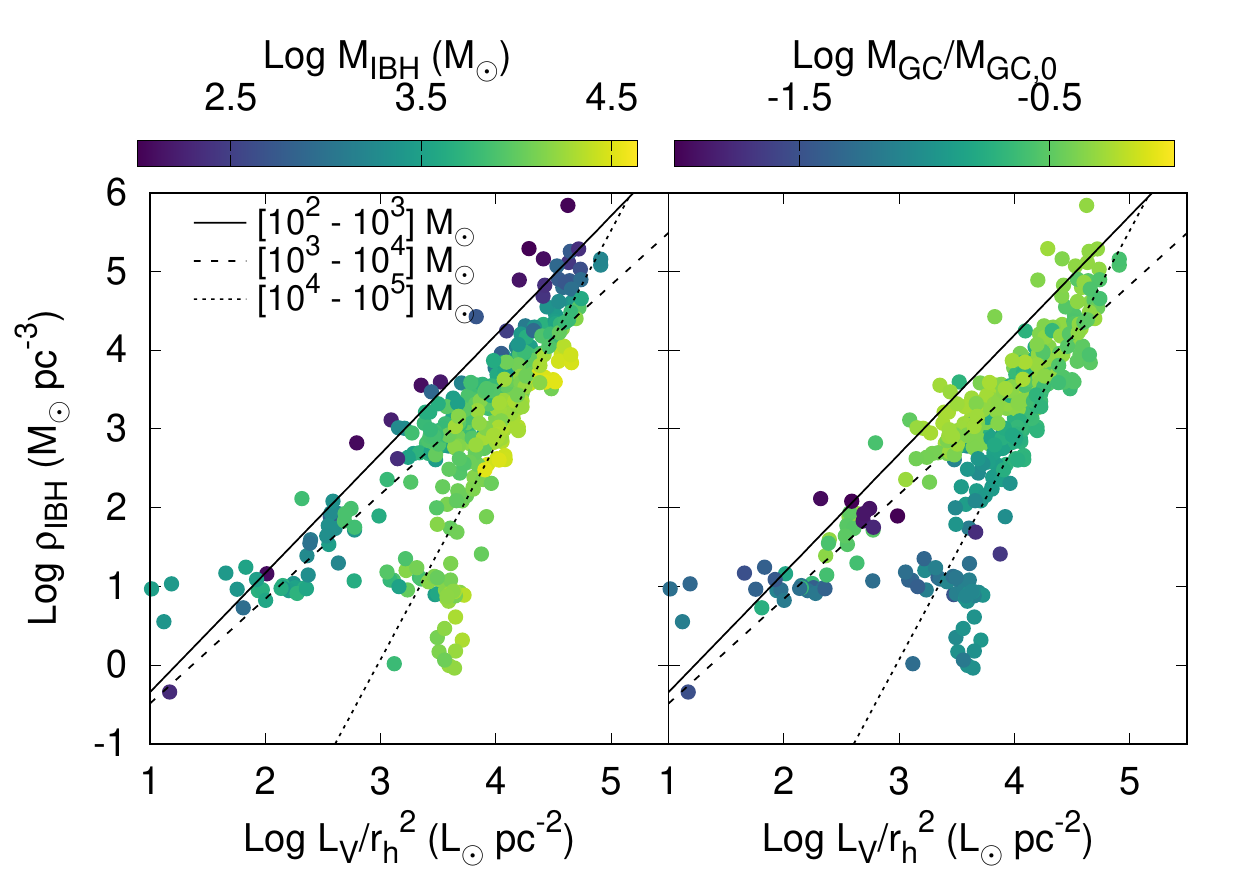}
    \caption{IMBHs fundamental plane: IMBH density as a function of the total visual luminosity divided by the square of the cluster observational half-mass radius. Different lines represent the best-fitting for models with IMBH masses in the range $10^2<M_\ibh/\Ms<10^3$ (straight line), $10^4<M_\ibh/\Ms<10^5$ (dashed line), and $10^4<M_\ibh/\Ms<10^5$ (dotted line). Color-coding identifies the IMBH mass (left panel) or the ratio between final and initial values of the GC mass. Final quantities are taken at 12 Gyr.}
    \label{fig:fundamental}
\end{figure*}

As introduced by \cite{AAG18a}, it is possible for IMBHs to define a fundamental plane, similarly to stellar BH subsystems, delimited on a side by the IMBH average density and, on the other side, by the host cluster average surface luminosity, defined as the ratio between the total visual luminosity ($L_{\rm V}$) and the squared observational half-mass radius ($r_h$)\footnote{Throughout the text we use observational half-mass radius and half-light radius indifferently}. Figure \ref{fig:fundamental} shows the fundamental plane for all the IMBHs found in MOCCA.
Intringuingly, we find that the link between the GC observational properties and the IMBH density depends critically on $M_\ibh$. As shown in left panel of Figure \ref{fig:fundamental}, gathering the MOCCA GCs population in three distinct samples that differ in the IMBH mass range. Each sample shows a tight relation, which is steeper for the heaviest IMBHs, $M_\ibh>10^4$. The fundamental relation is well fitted by a power-law, with form
\begin{equation}
\Log \rho_\ibh = \alpha\Log L_{\rm V}/r_h^2 + \beta,
\end{equation}
being the best-fit parameters $\alpha = 2.7\pm0.2$ and $\beta -8.1\pm 0.6$ for the heaviest IMBHs and $\alpha=1.33\pm 0.04$ and $\beta= -1.82\pm0.15$ for IMBHs in the mass range $10^3-10^4\Ms$.
At a fixed $L_{\rm V}$ value, heavier IMBHs are embedded in looser spheres of influence. Fixing $\rho_\ibh$, instead, makes apparent that heavier IMBHs are hosted in more luminous clusters. The right panel in Figure \ref{fig:fundamental} allows us to clarify why different IMBH mass ranges lead to a different arrangement in the fundamental plane. Indeed, the steep relation valid for 
the heaviest IMBHs is populated by clusters that lost more than $70\%$ of their initial mass. The fundamental relation for these ``disrupting'' clusters is much steeper compared to ``surviving'' clusters. This is mostly due to the fact that surviving clusters have core radii ranging in a relatively large range of values $1-10$ pc, while their $L_{\rm V}$ values are all peaked around $10^5\Ls$, leading the average surface luminosity $L_{\rm V}/r_h^2$ to increase proportionally to the square of the observational half-mass radius. Conversely, disrupting clusters have observational half-mass radii in a narrow range $1-2$ pc and luminosities spanning a large range, thus the increase of surface average luminosity is regulated by the linear dependence on $L_{\rm V}$. This can explain the different trends observed in the fundamental plane. 

Also the IMBH formation channels play a role in shaping the fundamental plane. Indeed, SLOW IMBHs tend to follow the correlation valid for surviving clusters, while FAST IMBHs lie on both the correlations. This simply reflects the fact that, in general, SLOW IMBHs live in GCs that preserve quite a large fraction of their initial mass, while a noticeable number of FAST IMBHs develop in GCs that undergo a severe mass loss over a Hubble time.

While the fundamental plane provides a useful tool to directly connect GCs observables with the IMBH, it does not allow to uniquely target potential IMBHs host candidates. Indeed as shown in our companion papers, GCs with similar properties might harbour an IMBH, a subsystem of stellar mass BHs, or be substantially {\it BH-free} \citep{AAG18a,AAG18b}. 

One possibility is that GCs hosting different populations (an IMBH, a BH subsystem, or simply stars) arrange differently in the plane defined by different observables. Unfortunately, determining what observational parameters maximize the differences is not an easy task. 

\begin{figure}
    \includegraphics[width=\columnwidth]{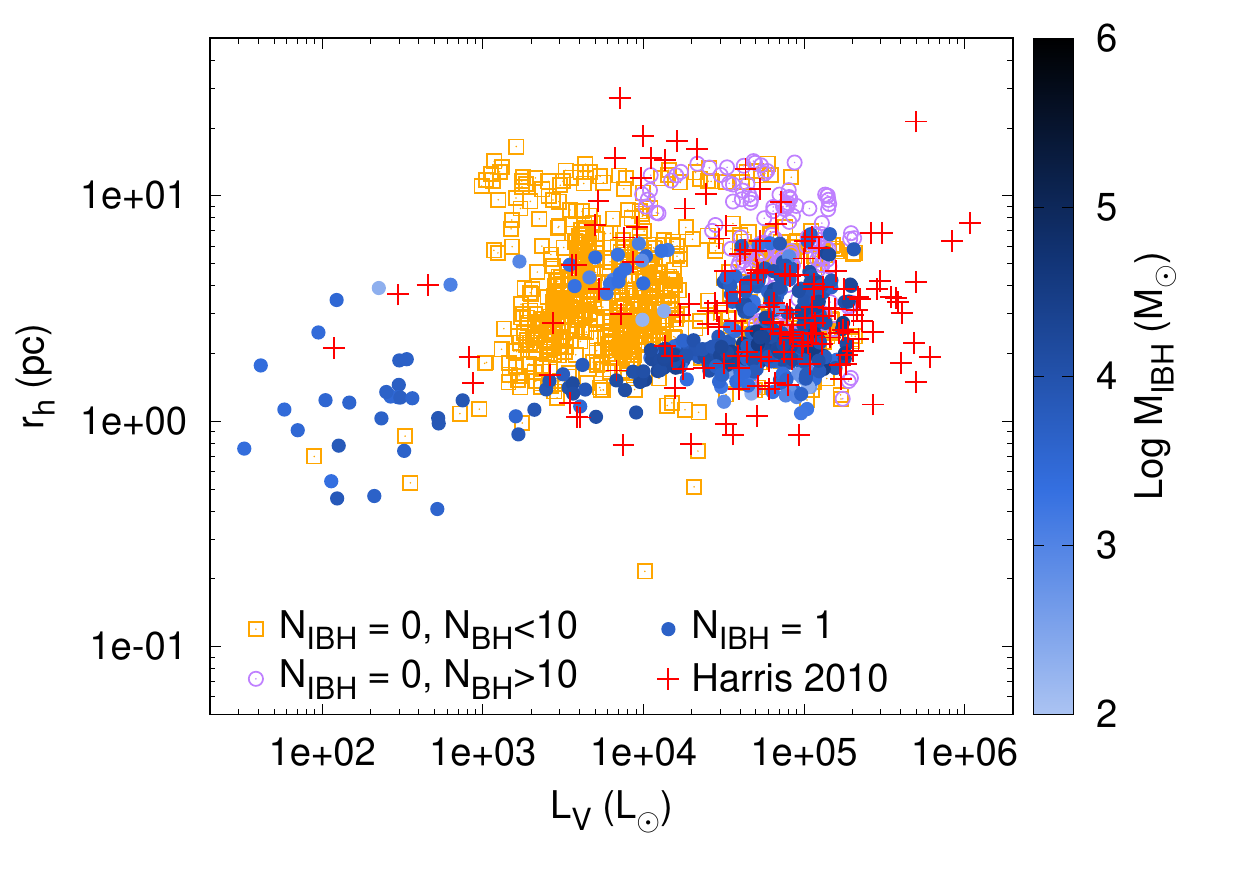}
    \caption{Observational half-mass radius as a function of the total visual luminosity for GCs in the \citet{harris10} catalogue (red crosses) and for MOCCA models containing, at 12 Gyr, either no BHs (open dgreen squares), at least 10 BHs with mass below $150\Ms$ (open purple points), or an IMBH (filled points). Colour-coding marks the IMBH mass, if present.
    }
    \label{fig:lvsrh}
\end{figure}

\begin{figure}
    \centering
    \includegraphics[width=\columnwidth]{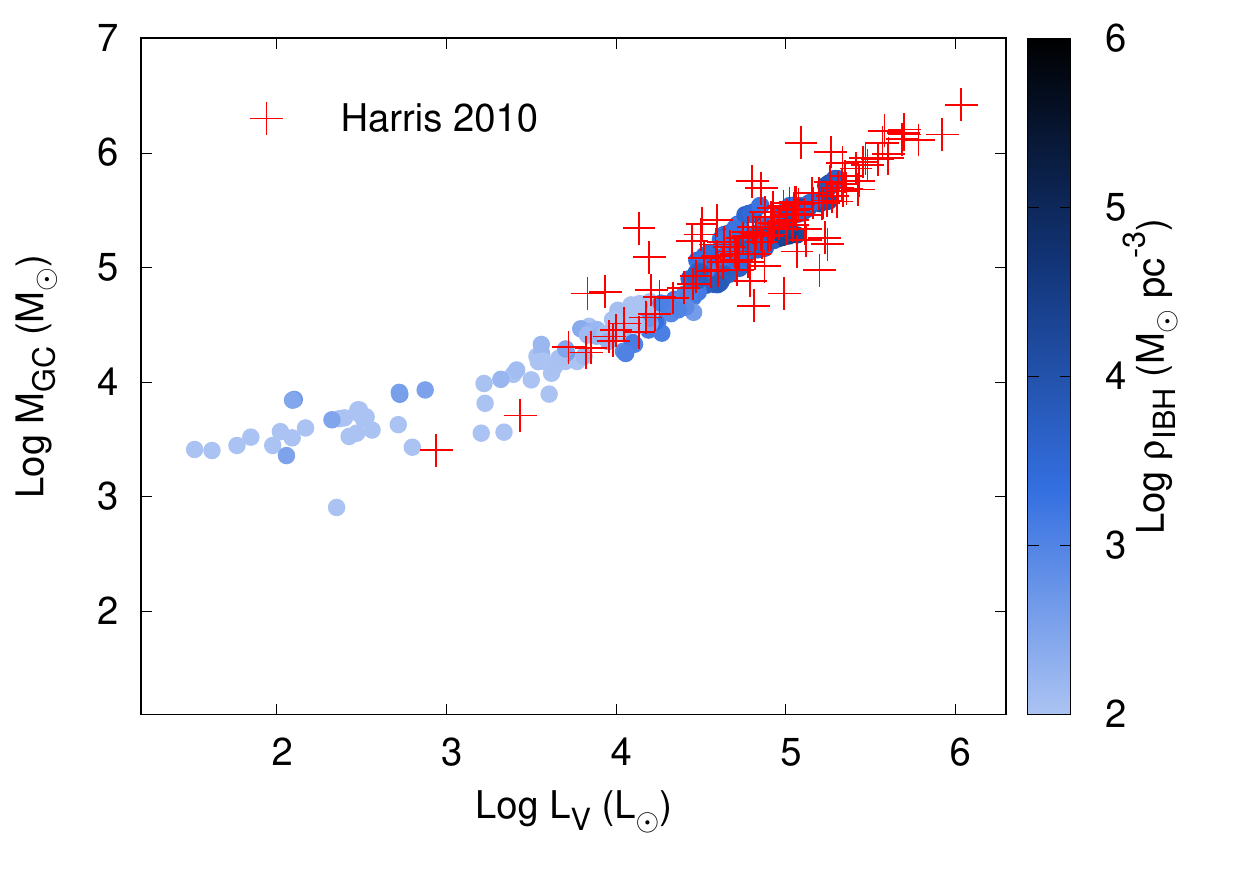}
    \caption{Present-day mass as a function of the total visual luminosity for Galactic GCs \citet{harris10} (red crosses) and MOCCA models that contain an IMBH (coloured points). The color map identifies the influence sphere average density.
    }
    \label{fig:comp}
\end{figure}

Figure \ref{fig:lvsrh} shows, for instance, how all MOCCA GCs distribute in the plane defined by half-mass radius and total visual luminosity, compared to actual Galactic GCs \citep[as taken from][catalogue]{harris10}. Broadly speaking, our models gather in three distinct sectors of the plane, with IMBH-dominated GCs that occupy the region of large luminosities and moderate $r_h$ values ($1-5$ pc). Models rich in stellar BHs occupy the same luminosity range, but are characterised by larger half-mass radius. Clusters that do not contain any appreciable BH population, instead, are characterised by a lower luminosity and a wide range of $r_h$ values. Although the separation between these models is quite straightforward, it must be noted a considerable overlapping between all of them, especially in the region $L_{V} = 10^4-10^5\Ls$ and $r_h = 1-10$ pc. Unfortunately, this is the region where most of Galactic GCs lie.  Another possible connection can be established between GCs luminosity and masses, as shown in Figure \ref{fig:comp}. The $M_{\gc}-L_{\rm V}$ plane is particularly interesting, as it shows a peculiar behaviour for GCs harbouring IMBHs surrounded by loose spheres of influence. MOCCA models having $\rho\gtrsim 3\times10^3$, as well as models containing a BH subsystem or stars-only, nicely distribute in the plane following a linear relation, as expected from simply converting the stellar mass in visual luminosity and vice-versa. Clusters with low-density spheres of influence, instead, significantly deviate from such relation, being characterized by a flatter distribution. Therefore, this kind of clusters appear less luminous than those with a similar mass and containing no-IMBH. Unfortunately, a direct comparison with observation is quite difficult owing to the intrinsecally different methods used to calculate the luminosity from the mass (in theoretical models) and the mass from the luminosity (in observations). 

\begin{figure}
    \centering
    \includegraphics[width=\columnwidth]{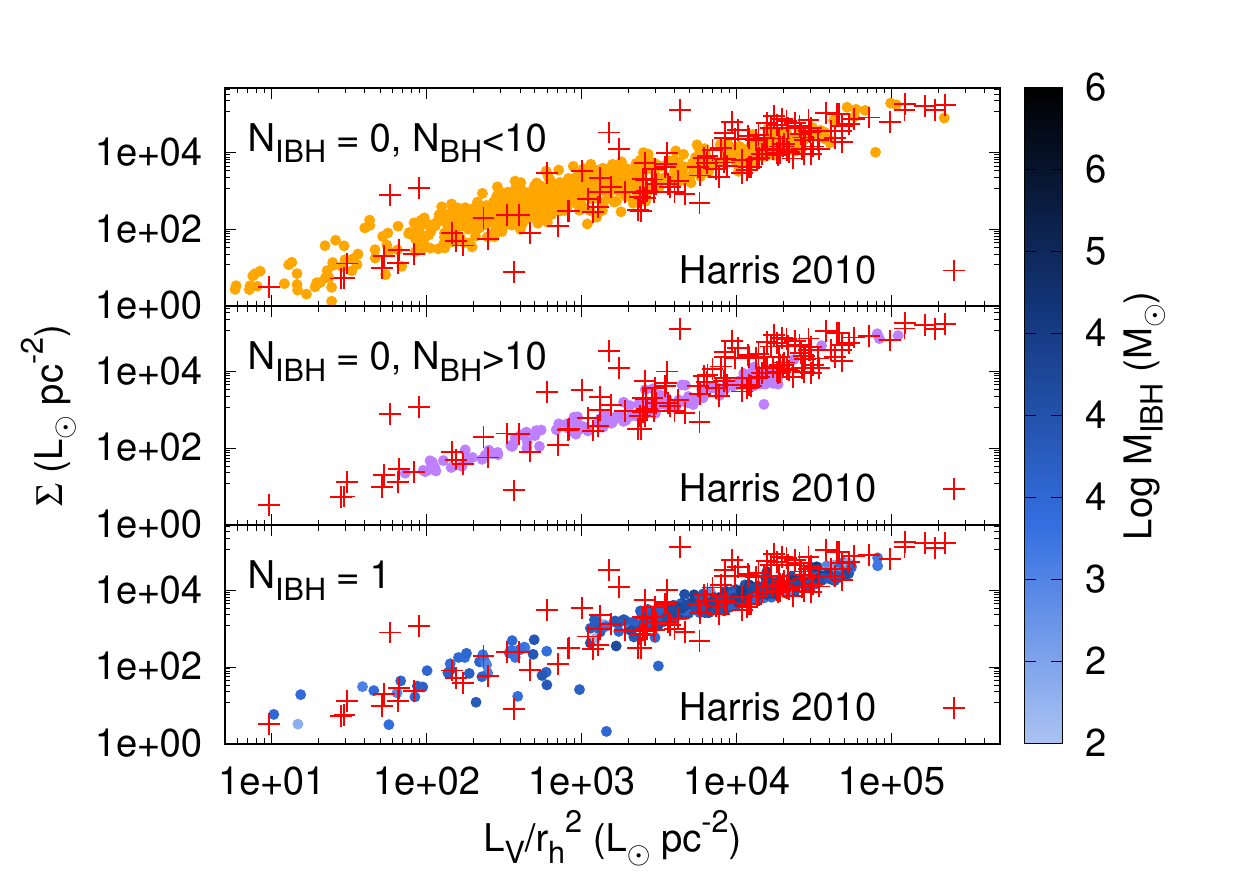}
    \caption{Central surface brightness as a function of the average surface brightness for Galactic GCs \citet{harris10} (red crosses) and for MOCCA models. Top panel shows only MOCCA models that have less than 10 BHs and no IMBH at 12 Gyr. Central panel shows only models having at least 10 BHs. Bottom panel shows models harbouring an IMBH. Color coding identifies the IMBH mass, if present, or the mass of the heaviest BH.
    }
    \label{fig:obs}
\end{figure}

Another possibility is to use only observed quantities, so to limit possible biases affecting their conversion into dynamical quantities. Figure \ref{fig:obs} shows GCs central surface brightness $\Sigma$ as a function of the average brightness used to define the fundamental plane, $L_{\rm V}/r_h^2$. We divided MOCCA models into those containing i) only a few BHs, ii) a BH subsystem, or iii) an IMBH, in order to better outline differences and similarities. The overlap between different models and Milky Way GCs is apparent, as well as the overlap between MOCCA models themselves. The inability to clearly distinguish between different models make hard to place constraints on possible IMBH-host candidates.
However, as we will show in the next section, this can be done if a large set of observed quantities are taken into account simultaneously. 

\section{A multi-dimensional method to discern GCs hosting IMBHs and BHS}
\label{norm}
In order to identify what Galactic GCs are most likely hosting at present time a BHS or an IMBH, we use a technique that relies upon the minimization of the differences between observed and simulated quantities. 

Indeed, MOCCA models aim at providing an overall representation of the Milky Way GC system, but they are not tailored to reproduce any of the observed Galactic GCs, thus a one-to-one correspondence between models and observations is unfeasible.

Therefore, to select MOCCA models that best reproduce a given GCs we calculate the norm
\begin{equation}
\norm{\mathcal{N}} = \sqrt{\left(\Sigma_i \left(\Delta \mathcal{V}_i\right)^2 \right) },
\end{equation}
being $\Delta \mathcal{V}_i$ the difference between the generic observable and the corresponding simulated quantity.
For each Galactic GC, we identify the 10 nearest MOCCA models, i. e. with the smallest norm. We target GCs as BHS- or IMBH-host if at least 5 among the 10 closest MOCCA models harbour either more than 10 BHs (BHS-dominated) or a BH with mass $>150\Ms$ (IMBH-dominated).

We calculate a 7-dimensional norm with components the GCs average surface brightness $L_{\rm V}/r_h^2$ and central surface brightness $\Sigma$, the observational core radius, $r_c$\footnote{In order to calculate the observational core radius, we use the technique described in Morscher et al. 2015 \citep[see Appendix in ][]{morscher15} that fits the \cite{king62} model (see Equation 13 in \citet{king62} or equation A2 in \citet{morscher15}) to the cumulative luminosity
of the cluster as a function of distance from the cluster centre. This binning-free technique obtains central surface
brightness and core radius values that avoid random noise from bright stars that can be close to the IMBH and can
drive up central surface brightness values.}, and half-light radius, $r_h$, the GCs Galactocentric distance, the visual $L_{\rm V}$ and bolometric luminosity $L_B$. All quantities are taken at 12 Gyr. Table \ref{longT} summarizes the results of the targeting procedure for all Galactic GCs\footnote{Note that the total number in this table is smaller than the actual number of Milky Way GCs due to the constraints used in our selection procedure.}.
For each GC, we use the scaling relations described here and in our companion paper \citep{AAG18a} to infer either the number of BHs and the BHS or the IMBH mass.

\begin{table*}
    \centering
    \caption{Number of putative IMBH and BHS in Galactic GCs}
    \begin{center}
        \begin{tabular}{ccccccc}
        \hline
        \hline
         $N_{DIM}$ & parameters & $N_{\rm IBH}$ & $N_{\rm BHS}$ & $N_{\rm none}$ & $N_{\rm UNC}$ & $N_{\rm tot}$ \\
        \hline
         7D & $L_{V}/r_h^2,~\Sigma, L_{V},L_{B},r_h, r_c, R_{\gc}$& 35 & 23 & 59&  19 & 136\\
        \hline
        \end{tabular}
    \end{center}
    \begin{tablenotes}
        \item Col. 1: number of dimensions used to define the norm. Col. 2: quantities used to calculate the norm. Col. 3(4): Number of Galactic GCs containing an IMBH(BHS). Col. 5-6: Number of Galactic GCs that does not contain any central dark object, or for which a clear assesment cannot be made. Col 7: total number of GCs.
        \end{tablenotes}
    \label{tab:t3}
\end{table*}

In five cases, namely Whiting1, Pal1, Ko1, Ko2, and AM4, the GC luminosity has values below $\sim 3\times 10^3\Ls$. In our MOCCA database, GCs containing an IMBH and having such low luminosities are characterized by influence radius almost constant $R_\ibh \simeq 3.6-10$ pc. In this case, we used directly the $R_\ibh-M_\ibh$ relation, rather than using the density of the sphere of influence. Nonetheless, we note that the correlations predict very massive IMBHs compared to the GC observed mass and luminosity. Moreover, we show in previous sections that correlations tend to be much looser at GCs current masses below $\sim 1.5\times 10^4\Ms$. Therefore, in the following we only consider GCs with a present day mass above $1.5\times 10^4\Ms$ and visual luminosity $>3\times 10^3\Ls$.

Under this approximation and assuming the 7D norm, we find $35$ GCs that might be harbouring an IMBH at present and $23$ potentially containing a BHS. Models containing a BHS in this selection overlap pretty well with our previous paper \citep{AAG18b}, although the matching is not $100\%$. For 19 clusters we cannot say whether they host an IMBH or a BHS, while for the remaining 59 GCs our analysis do not suggest any significant central massive object. All these quantities are summarized in Table \ref{tab:t3}.

\begin{figure}
    \centering
    \includegraphics[width=\columnwidth]{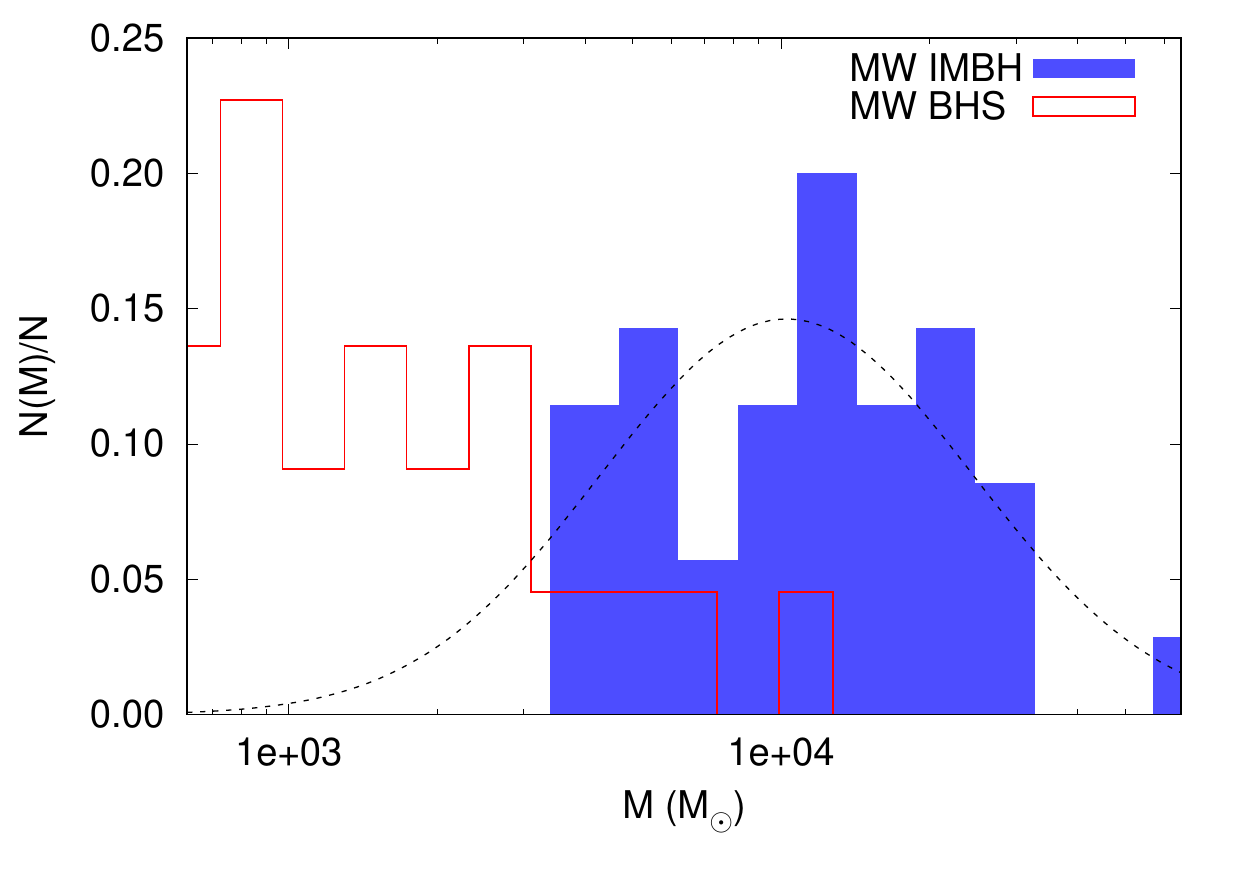}
    \caption{Mass distribution of central object for Galactic GCs containing an IMBH (blue filled steps), or a BHS (red open steps). }
    \label{fig:MWDARK}
\end{figure}

We show in Figure \ref{fig:MWDARK} the mass distribution of BHSs and IMBHs inferred for MW clusters. Despite the low number statistics, we find that the distribution of logarithmic IMBH mass can be described by a Gaussian 
\begin{equation}
    f(M_\ibh) = \frac{a}{\sigma\sqrt(2\pi)}\exp{\left[-\frac{\left(\Log (M_\ibh/\Ms ) - \mu\right)^2}{2\sigma^2}\right]},
\end{equation}
with $a=0.14\pm0.3$ , $\mu = 4.01\pm0.09$ and $\sigma = 0.4\pm0.1 $. Most of IMBHs masses inferred with our approach exceed $10^3\Ms$.

\begin{figure}
    \centering
    \includegraphics[width=\columnwidth]{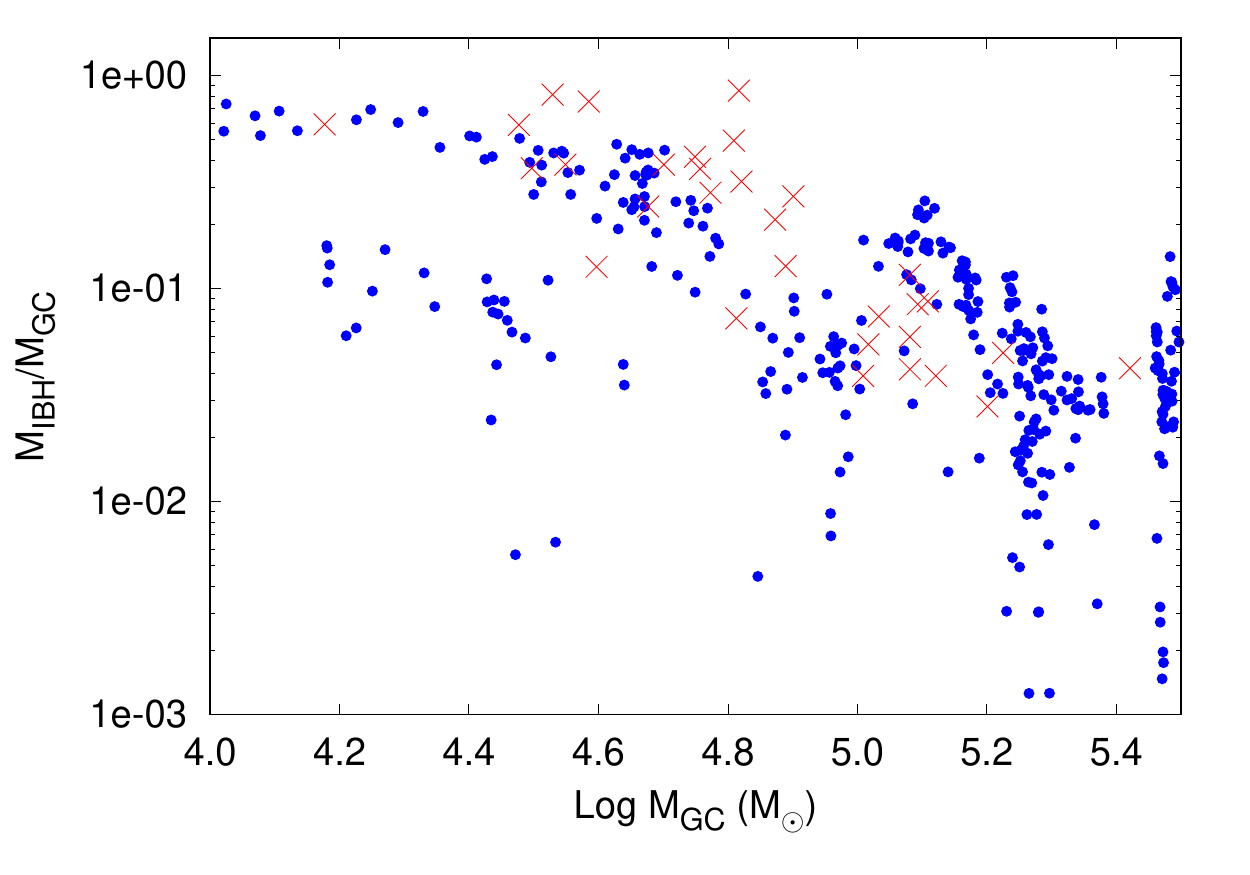}
    \caption{IMBH-to-GC mass ratio for MOCCA models (blue filled points) and for shortlisted Galactic GC (red crosses).}
    \label{fig:ratio}
\end{figure}

Using MOCCA models, we find that the ratio $M_\ibh/M_{\gc}$ decreases at increasing the GC mass, as shown in Figure \ref{fig:ratio}. Clearly, this relation affect the IMBH mass inferred for Galactic GCs. Indeed, out of the 35 shortlisted GCs, we find that 19 are characterized by $M_{\ibh}/M_{\gc}>0.1$, 6 have $0.05<M_{\ibh}/M_{\gc}<0.1$, and the remaining 10 have $M_{\ibh}/M_{\gc}<0.05$. 

Possible GCs hosting an IMBH might be NGC1851 ($ M_\ibh \simeq 3.48\times 10^3 \Ms$), NGC6093 ($ M_\ibh \simeq 3.63\times 10^3 \Ms$), NGC6254 ($ M_\ibh \simeq 8.39\times 10^3 \Ms$). On the other hand GCs possibly harboring a noticeable number of BHs are NGC 3201 $N_\bh \sim 113$ and NGC 6101 $N_\bh \sim 147$, which are already known in literature as possible site with a large BH number.

The heaviest IMBH found through our analysis is hosted in NGC 6558, with an inferred mass of $M_\ibh = 3.2\times 10^4\Ms$, similar to the value inferred from microlensing measurements provided by \cite{safonova10}, in NGC 6681 ($M_\ibh \simeq 7.2\times 10^3\Ms$), and NGC 6397 ($M_\ibh \simeq 9.9\times 10^3\Ms$). 
Moreover, the procedure described here has the potential to be used for excluding the presence of an IMBH in GC. For instance, we found that cluster NGC 288 and NGC 5466 more likely host a BHS as massive as $2520\Ms$ and $4170\Ms$, rather than an IMBH, as suggested by earlier works \citep{lutz13}.

Our approach provides a rapid way to shortlist GCs which may be of potential interests for tailored numerical studies or further observational campaigns.
However, we must note that the core radius might be ill-defined in presence of a central IMBH, while the typical luminosity and or central surface brightness can be dominated by the brightest stars in absence of a BHS, possibly causing misleading correspondence between models and observations. Therefore, we caution that a more careful analysis must be performed to assess whether a GC hosts an IMBH or a BHS.

\section{Conclusion}
\label{end}

In this paper, we use the MOCCA SURVEY DATABASE I to dissect the interplay between IMBHs and their nursing GCs. 
Making use of over 2000 GC models, we show that it is possible to define a fundamental plane connecting the IMBH sphere of influence and the host GC luminosity and core radius.
Our main results can be summarized as follows:
\begin{itemize}
\item we provide an extensive analysis of 407 MOCCA models containing an IMBH at 12 Gyr, characterizing each IMBH and its surrounding through the IMBH mass, influence radius, stellar density enclosed within the influence radius, and formation time. 
\item we show that IMBHs forming in MOCCA have typical masses $\sim 10^3-10^4\Ms$ characterized by an influence radii distribution peaking at $\sim 3$ pc, as shown in top and central panels of Figure \ref{fig:formation};
\item IMBHs formation time shows a clear bimodality that allows us to distinguish FAST (formation time smaller than 1 Gyr) and SLOW IMBHs. FAST IMBHs are, on average, heavier and surrounded by lower-mass stars compared to SLOW IMBHs (see bottom panel of Figure \ref{fig:formation};
\item we find that IMBH interactions with surrounding GC members drive the formation of a binary system in $32\%$ of the cases, as shown in Figure \ref{fig:IBHinBin}, where the IMBH is bound to a MS star ($13.7\%$), a WD ($16.2\%$), a NS ($0.2\%$), or a BH ($2.7\%$). These systems are extremely important, as they can trigger tidal disruption events (in the case of MS), and the formation of intermediate mass ratio inspirals emitting GWs (for BHs) possibly in combination with explosive events and electromagnetic emission (for WDs and NSs). We list the main properties of GW sources candidates in Table \ref{tab:catalogue};
\item we find a striking correlation between IMBH mass and GC central density at 12 Gyr, provided that the GC mass at present time is above $5\times 10^4\Ms$, as shown in Figure \ref{fig:gcrelibh};
\item we discover a population of {\it dark clusters}, i.e. clusters losing more than $90\%$ of their mass, characterized by orbits within 3-5 kpc from the Galactic Centre. These class of clusters contain, on average, IMBHs with masses above $5\times 10^3\Ms$, which constitutes more than $10\%$ of the whole GC mass at 12 Gyr. Due to their low-orbits, these clusters can reach the Galactic Centre within a Hubble time, possibly dragging in there their IMBHs. These would nicely explain several recent observations of putative IMBHs orbiting the inner 10 pc of the Milky Way centre (see Figure \ref{fig:ibhmult});
\item IMBHs forming via dynamical interactions in MOCCA can provide clues on the origin of SMBHs. We show that a fitting formula applied to MOCCA models and extended to the mass range of dwarf galaxies nicely overlap with a sample of observed low-mass AGN in dwarf galaxies (see Figure \ref{fig:mscaling}); 
\item we find a {\it fundamental plane} linking IMBH sphere of influence density and the host GC total luminosity and observational half-mass radius. We find two clear sequences in the fundamental plane. One sequence is defined by IMBHs with masses $<10^4\Ms$ residing in GCs preserving more than $30\%$ of their initial mass. The second sequence, instead, is dominated by heavier IMBHs, residing in clusters which underwent a severe mass loss, being $M_{\gc}/M_\gc<0.3$. The fundamental plane is shown in Figure \ref{fig:fundamental};
\item we propose a simple tool to target Milky Way GCs possibly harbouring an IMBH at present. This concept relies upon the definition of a 7 dimensional space defined by GCs visual and bolometric total luminosity, half-mass and core radii, galactocentric distance, average and central surface luminosity. In this 7D space, for each Galactic GCs we find the 10 closest MOCCA models and calculate how many of them contain an IMBH, a BHS, or nothing. We shortlist 35(23) Galactic GCs possibly harbouring an IMBH(BHS) at present time, as summarized in Tables \ref{tab:t3} and \ref{longT}.
\end{itemize}

Our ranking procedure provides a simple and rapid tool to identify GCs hosting a central {\it dark} object (either an IMBH or a BH subsystem) and might be used to shortlist potential candidates to be observed with tailored observations.

In future works we will try to better constrain the discussed in the paper correlations using the new MOCCA models with updated physics and better determination of cluster parameters in the case of an IMBH presence.

\section{Acknowledgments}

MAS also acknowledges financial support from the Alexander von Humboldt Foundation and the Federal Ministry for Education and Research in the framework of the research project ``The evolution of black holes from stellar to galactic scales'', and the Sonderforschungsbereich SFB 881 ``The Milky Way System'' of the German Research Foundation (DFG). MAS acknowledges the Nicolaus Copernicus Astronomical Center in Warsaw for the hospitality given during the development of part of this research. MG was partially supported by the Polish National Science Center (NCN)  through the grant UMO-2016/23/B/ST9/02732. AA is currently supported by the Carl Tryggers Foundation for Scientific Research through the grant CTS 17:113.  This work benefited from support by the International Space Science In-stitute (ISSI), Bern, Switzerland, through its International Team programme ref. no. 393 The Evolution of Rich Stellar Populations \& BH Binaries (2017-18).

\clearpage
\footnotesize{
\bibliographystyle{mnras}
\bibliography{biblio}
}

\appendix
\section{A}

In Table \ref{longT} we show, for all the Galactic GCs, the probability to host an IMBH or a BHS assuming a 7D norm as explained in Section \ref{norm}. Upon the definitions assumed here, we provide a label indicating whether the GC is likely to host an ``IMBH'', a ``BHS'' or ``NONE'', an estimate of BHs number (1 for IMBH), and the mass of the central dark object, either the IMBH or the BHS. For all target identified in \cite{AAG18b}, for BHS, in the table we mark in red those that we find also with our new approach, in green those that gives an unclear results, in blue those that according to our approach should contain an IMBH, and in purple those that according to our approach should contain neither a BHS nor an IMBH.

\onecolumn
\small
\setlength\LTleft{0pt}
\setlength\LTright{0pt}
\LTcapwidth=\textwidth
\begin{longtable}{ccccccccc}
\caption{Galactic GCs probability to host an IMBH. }
\label{longT}
\\
\multicolumn{8}{l}{Red color marks GCs with BHS find also in \citet{AAG18b}, green marks those giving unclear results,} & \\
\multicolumn{8}{l}{blue marks GCs in \citet{AAG18b} sample potentially harbouring an IMBH according to the new approach, } & \\
\multicolumn{8}{l}{purple marks those having no central dark object (IMBH or BHS).} &
\\
\hline
\multirow{3}{*}{ID} & \multicolumn{4}{c}{7D norm} & & & & \\
   &  $f_{\rm none}$ & $f_\bhs$ & $f_\ibh$ & central  & $N_{\bh}$ & $M_{\rm cen}$& $M_{\rm obs}$ & $L_{\rm V,obs}$ \\
   & $\%$ & $\%$ & $\%$ &object & &$10^3~\Ms$ &$10^5~\Ms$&$10^5~\Ls$\\
\hline
\endfirsthead
\multicolumn{8}{c}%
{series \tablename\ \thetable{} -- continued from previous page} \\
\hline \multirow{3}{*}{ID} & \multicolumn{4}{c}{7D norm}  & & & \\
   & $f_{\rm none}$ & $f_\bhs$ & $f_\ibh$ & central  & $N_{\bh}$ & $M_{\rm cen}$ & $M_{\rm obs}$ & $L_{\rm V,obs}$\\
   & $\%$ & $\%$ & $\%$ & object& &$10^3~\Ms$ &$10^5~\Ms$ &$10^5~\Ms$ \\
\hline
\endhead
\hline \multicolumn{9}{r}{{Continued on next page}} \\ \hline
\endfoot
\endlastfoot
NGC104	& $80.0 $&$ 0.0 $&$ 20.0 $& NONE &  -  &  - &$10.02$ &$5.01$  \\
{\color{red} {\bf NGC288}}	& $9.1 $&$ 90.9 $&$ 0.0 $& BHS & $191 \pm 41$ & $2.52\pm 0.57$ &$0.86$ &$0.43$  \\
NGC362	& $58.2 $&$ 0.0 $&$ 41.8 $& NONE &  -  &  - &$4.03$ &$2.01$  \\
Whiting1	& $0.0 $&$ 0.0 $&$ 100.0 $& IMBH &  1  & $67.81\pm 19.02$ &$0.02$ &$0.01$  \\
NGC1261	& $1.8 $&$ 52.7 $&$ 45.5 $& BHS & $74 \pm 16$ & $0.88\pm 0.20$ &$2.25$ &$1.13$  \\
Pal1	& $0.0 $&$ 0.0 $&$ 100.0 $& IMBH &  1  & $38.36\pm 10.43$ &$0.02$ &$0.01$  \\
AM1	& $100.0 $&$ 0.0 $&$ 0.0 $& NONE &  -  &  - &$0.13$ &$0.07$  \\
Eridanus	& $65.5 $&$ 34.5 $&$ 0.0 $& NONE &  -  &  - &$0.19$ &$0.10$  \\
Pal2	& $0.0 $&$ 43.6 $&$ 56.4 $& IMBH &  1  & $11.17\pm 2.64$ &$2.64$ &$1.32$  \\
NGC1851	& $20.0 $&$ 16.4 $&$ 63.6 $& IMBH &  1  & $3.48\pm 0.91$ &$3.67$ &$1.84$  \\
NGC1904	& $94.5 $&$ 5.5 $&$ 0.0 $& NONE &  -  &  - &$2.38$ &$1.19$  \\
NGC2298	& $40.0 $&$ 0.0 $&$ 60.0 $& IMBH &  1  & $20.91\pm 4.80$ &$0.57$ &$0.29$  \\
NGC2419	& $0.0 $&$ 100.0 $&$ 0.0 $& BHS & $205 \pm 45$ & $2.73\pm 0.62$ &$10.02$ &$5.01$  \\
Ko2	& $0.0 $&$ 0.0 $&$ 100.0 $& IMBH &  1  & $63.04\pm 17.60$ &$0.00$ &$0.00$  \\
NGC2808	& $56.4 $&$ 23.6 $&$ 20.0 $& NONE &  -  &  - &$9.75$ &$4.88$  \\
E3	& $0.0 $&$ 0.0 $&$ 100.0 $& IMBH &  1  & $127.68\pm 30.78$ &$0.08$ &$0.04$  \\
Pal3	& $67.3 $&$ 32.7 $&$ 0.0 $& NONE &  -  &  - &$0.32$ &$0.16$  \\
{\color{red} {\bf NGC3201}}	& $0.0 $&$ 61.8 $&$ 38.2 $& BHS & $113 \pm 24$ & $1.41\pm 0.31$ &$1.63$ &$0.82$  \\
Pal4	& $49.1 $&$ 50.9 $&$ 0.0 $& BHS & $645 \pm 175$ & $9.70\pm 2.77$ &$0.43$ &$0.22$  \\
Ko1	& $0.0 $&$ 0.0 $&$ 100.0 $& IMBH &  1  & $64.29\pm 17.97$ &$0.01$ &$0.00$  \\
NGC4147	& $34.5 $&$ 0.0 $&$ 65.5 $& IMBH &  1  & $19.23\pm 4.42$ &$0.50$ &$0.25$  \\
{\color{red} {\bf NGC4372}}	& $7.3 $&$ 92.7 $&$ 0.0 $& BHS & $140 \pm 29$ & $1.79\pm 0.40$ &$2.23$ &$1.12$  \\
Rup106	& $74.5 $&$ 16.4 $&$ 9.1 $& NONE &  -  &  - &$0.59$ &$0.30$  \\
{\color{dgreen} {\bf NGC4590}}	& $20.0 $&$ 38.2 $&$ 41.8 $& UNCLEAR &  -  &  - &$1.52$ &$0.76$  \\
{\color{blue} {\bf NGC4833}}	& $9.1 $&$ 5.5 $&$ 85.5 $& IMBH &  1  & $12.09\pm 2.84$ &$3.17$ &$1.58$  \\
NGC5024	& $49.1 $&$ 21.8 $&$ 29.1 $& UNCLEAR &  -  &  - &$5.21$ &$2.61$  \\
NGC5053	& $30.9 $&$ 69.1 $&$ 0.0 $& BHS & $397 \pm 97$ & $5.67\pm 1.44$ &$0.87$ &$0.43$  \\
NGC5139	& $7.3 $&$ 65.5 $&$ 27.3 $& BHS & $57 \pm 13$ & $0.67\pm 0.16$ &$0.00$ &$10.86$  \\
{\color{blue} {\bf NGC5272}}	& $32.7 $&$ 16.4 $&$ 50.9 $& IMBH &  1  & $13.10\pm 3.07$ &$6.10$ &$3.05$  \\
NGC5286	& $92.7 $&$ 0.0 $&$ 7.3 $& NONE &  -  &  - &$5.36$ &$2.68$  \\
AM4	& $20.0 $&$ 0.0 $&$ 80.0 $& IMBH &  1  & $74.30\pm 20.96$ &$0.01$ &$0.00$  \\
{\color{red} {\bf NGC5466}}	& $27.3 $&$ 72.7 $&$ 0.0 $& BHS & $301 \pm 69$ & $4.17\pm 1.00$ &$1.06$ &$0.53$  \\
NGC5634	& $80.0 $&$ 9.1 $&$ 10.9 $& NONE &  -  &  - &$2.04$ &$1.02$  \\
NGC5694	& $61.8 $&$ 0.0 $&$ 38.2 $& NONE &  -  &  - &$2.32$ &$1.16$  \\
{\color{red} {\bf IC4499}}	& $0.0 $&$ 100.0 $&$ 0.0 $& BHS & $232 \pm 51$ & $3.13\pm 0.72$ &$1.45$ &$0.72$  \\
NGC5824	& $100.0 $&$ 0.0 $&$ 0.0 $& NONE &  -  &  - &$5.93$ &$2.96$  \\
Pal5	& $100.0 $&$ 0.0 $&$ 0.0 $& NONE &  -  &  - &$0.20$ &$0.10$  \\
{\color{red} {\bf NGC5897}}	& $0.0 $&$ 100.0 $&$ 0.0 $& BHS & $197 \pm 43$ & $2.62\pm 0.59$ &$1.33$ &$0.67$  \\
NGC5904	& $32.7 $&$ 21.8 $&$ 45.5 $& UNCLEAR &  -  &  - &$5.72$ &$2.86$  \\
NGC5927	& $100.0 $&$ 0.0 $&$ 0.0 $& NONE &  -  &  - &$2.28$ &$1.14$  \\
NGC5946	& $96.4 $&$ 0.0 $&$ 3.6 $& NONE &  -  &  - &$1.27$ &$0.64$  \\
BH176	& $32.7 $&$ 0.0 $&$ 67.3 $& IMBH &  1  & $132.06\pm 31.92$ &$0.07$ &$0.04$  \\
{\color{red} {\bf NGC5986}}	& $9.1 $&$ 85.5 $&$ 5.5 $& BHS & $52 \pm 12$ & $0.61\pm 0.15$ &$4.06$ &$2.03$  \\
Lynga7	& $21.8 $&$ 0.0 $&$ 78.2 $& IMBH &  1  & $15.77\pm 3.66$ &$0.75$ &$0.37$  \\
Pal14	& $94.5 $&$ 0.0 $&$ 5.5 $& NONE &  -  &  - &$0.14$ &$0.07$  \\
NGC6093	& $27.3 $&$ 0.0 $&$ 72.7 $& IMBH &  1  & $3.63\pm 0.95$ &$3.35$ &$1.67$  \\
NGC6121	& $0.0 $&$ 0.0 $&$ 100.0 $& IMBH &  1  & $11.21\pm 2.65$ &$1.29$ &$0.64$  \\
{\color{red} {\bf NGC6101}}	& $14.5 $&$ 85.5 $&$ 0.0 $& BHS & $147 \pm 31$ & $1.89\pm 0.42$ &$1.02$ &$0.51$  \\
{\color{dgreen} {\bf NGC6144}}	& $30.9 $&$ 27.3 $&$ 41.8 $& UNCLEAR &  -  &  - &$0.94$ &$0.47$  \\
NGC6139	& $100.0 $&$ 0.0 $&$ 0.0 $& NONE &  -  &  - &$3.78$ &$1.89$  \\
Terzan3	& $47.3 $&$ 0.0 $&$ 52.7 $& IMBH &  1  & $46.44\pm 10.63$ &$0.14$ &$0.07$  \\
{\color{blue} {\bf NGC6171}}	& $0.0 $&$ 0.0 $&$ 100.0 $& IMBH &  1  & $14.01\pm 3.27$ &$1.21$ &$0.60$  \\
1636-283	& $1.8 $&$ 0.0 $&$ 98.2 $& IMBH &  1  & $24.16\pm 5.52$ &$0.07$ &$0.03$  \\
{\color{red} {\bf NGC6205}}	& $9.1 $&$ 54.5 $&$ 36.4 $& BHS & $58 \pm 13$ & $0.68\pm 0.16$ &$4.50$ &$2.25$  \\
NGC6229	& $14.5 $&$ 72.7 $&$ 12.7 $& BHS & $65 \pm 14$ & $0.78\pm 0.18$ &$2.86$ &$1.43$  \\
NGC6218	& $18.2 $&$ 61.8 $&$ 20.0 $& BHS & $71 \pm 15$ & $0.85\pm 0.20$ &$1.44$ &$0.72$  \\
FSR1735	& $30.9 $&$ 0.0 $&$ 69.1 $& IMBH &  1  & $4.72\pm 1.20$ &$0.65$ &$0.33$  \\
NGC6235	& $43.6 $&$ 0.0 $&$ 56.4 $& IMBH &  1  & $23.40\pm 5.35$ &$0.56$ &$0.28$  \\
NGC6254	& $0.0 $&$ 30.9 $&$ 69.1 $& IMBH &  1  & $8.39\pm 2.02$ &$1.68$ &$0.84$  \\
NGC6256	& $16.4 $&$ 0.0 $&$ 83.6 $& IMBH &  1  & $10.50\pm 2.49$ &$1.24$ &$0.62$  \\
Pal15	& $69.1 $&$ 30.9 $&$ 0.0 $& NONE &  -  &  - &$0.27$ &$0.14$  \\
NGC6266	& $43.6 $&$ 49.1 $&$ 7.3 $& UNCLEAR &  -  &  - &$8.04$ &$4.02$  \\
NGC6273	& $87.3 $&$ 7.3 $&$ 5.5 $& NONE &  -  &  - &$7.67$ &$3.84$  \\
NGC6284	& $100.0 $&$ 0.0 $&$ 0.0 $& NONE &  -  &  - &$2.61$ &$1.31$  \\
NGC6287	& $56.4 $&$ 18.2 $&$ 25.5 $& NONE &  -  &  - &$1.50$ &$0.75$  \\
NGC6293	& $60.0 $&$ 0.0 $&$ 40.0 $& NONE &  -  &  - &$2.21$ &$1.11$  \\
NGC6304	& $87.3 $&$ 0.0 $&$ 12.7 $& NONE &  -  &  - &$1.42$ &$0.71$  \\
NGC6316	& $100.0 $&$ 0.0 $&$ 0.0 $& NONE &  -  &  - &$3.71$ &$1.85$  \\
NGC6341	& $100.0 $&$ 0.0 $&$ 0.0 $& NONE &  -  &  - &$3.29$ &$1.64$  \\
NGC6325	& $14.5 $&$ 0.0 $&$ 85.5 $& IMBH &  1  & $5.69\pm 1.42$ &$1.04$ &$0.52$  \\
NGC6333	& $40.0 $&$ 21.8 $&$ 38.2 $& UNCLEAR &  -  &  - &$2.59$ &$1.29$  \\
NGC6342	& $65.5 $&$ 0.0 $&$ 34.5 $& NONE &  -  &  - &$0.63$ &$0.32$  \\
NGC6356	& $34.5 $&$ 29.1 $&$ 36.4 $& UNCLEAR &  -  &  - &$4.34$ &$2.17$  \\
NGC6355	& $54.5 $&$ 0.0 $&$ 45.5 $& NONE &  -  &  - &$2.89$ &$1.45$  \\
NGC6352	& $38.2 $&$ 0.0 $&$ 61.8 $& IMBH &  1  & $21.10\pm 4.84$ &$0.66$ &$0.33$  \\
IC1257	& $83.6 $&$ 0.0 $&$ 16.4 $& NONE &  -  &  - &$0.49$ &$0.25$  \\
Terzan2	& $21.8 $&$ 0.0 $&$ 78.2 $& IMBH &  1  & $29.15\pm 6.65$ &$0.38$ &$0.19$  \\
NGC6366	& $18.2 $&$ 0.0 $&$ 81.8 $& IMBH &  1  & $27.60\pm 6.30$ &$0.34$ &$0.17$  \\
Terzan4	& $12.7 $&$ 0.0 $&$ 87.3 $& IMBH &  1  & $77.39\pm 18.05$ &$0.11$ &$0.05$  \\
HP1	& $10.9 $&$ 0.0 $&$ 89.1 $& IMBH &  1  & $55.98\pm 12.87$ &$0.66$ &$0.33$  \\
{\color{dgreen} {\bf NGC6362}}	& $32.7 $&$ 40.0 $&$ 27.3 $& UNCLEAR &  -  &  - &$1.03$ &$0.52$  \\
NGC6380	& $85.5 $&$ 14.5 $&$ 0.0 $& NONE &  -  &  - &$1.71$ &$0.86$  \\
Terzan1	& $0.0 $&$ 0.0 $&$ 100.0 $& IMBH &  1  & $178.58\pm 44.30$ &$0.10$ &$0.05$  \\
Ton2	& $52.7 $&$ 0.0 $&$ 47.3 $& NONE &  -  &  - &$0.50$ &$0.25$  \\
NGC6388	& $45.5 $&$ 49.1 $&$ 5.5 $& UNCLEAR &  -  &  - &$9.93$ &$4.97$  \\
NGC6402	& $20.0 $&$ 50.9 $&$ 29.1 $& BHS & $46 \pm 10$ & $0.53\pm 0.13$ &$7.47$ &$3.73$  \\
{\color{purple} {\bf NGC6401}}	& $63.6 $&$ 10.9 $&$ 25.5 $& NONE &  -  &  - &$2.47$ &$1.24$  \\
NGC6397	& $12.7 $&$ 0.0 $&$ 87.3 $& IMBH &  1  & $9.89\pm 2.36$ &$0.77$ &$0.39$  \\
Pal6	& $52.7 $&$ 0.0 $&$ 47.3 $& NONE &  -  &  - &$0.89$ &$0.44$  \\
{\color{purple} {\bf NGC6426}}	& $69.1 $&$ 5.5 $&$ 25.5 $& NONE &  -  &  - &$0.80$ &$0.40$  \\
Djorg1	& $78.2 $&$ 1.8 $&$ 20.0 $& NONE &  -  &  - &$1.06$ &$0.53$  \\
Terzan5	& $1.8 $&$ 0.0 $&$ 98.2 $& IMBH &  1  & $4.45\pm 1.14$ &$1.59$ &$0.79$  \\
NGC6440	& $45.5 $&$ 49.1 $&$ 5.5 $& UNCLEAR &  -  &  - &$5.41$ &$2.70$  \\
NGC6441	& $49.1 $&$ 49.1 $&$ 1.8 $& UNCLEAR &  -  &  - &$12.16$ &$6.08$  \\
Terzan6	& $50.9 $&$ 0.0 $&$ 49.1 $& NONE &  -  &  - &$1.86$ &$0.93$  \\
NGC6453	& $45.5 $&$ 0.0 $&$ 54.5 $& IMBH &  1  & $5.15\pm 1.30$ &$1.32$ &$0.66$  \\
{\color{red} {\bf NGC6496}}	& $32.7 $&$ 67.3 $&$ 0.0 $& BHS & $98 \pm 20$ & $1.20\pm 0.27$ &$1.30$ &$0.65$  \\
Terzan9	& $0.0 $&$ 0.0 $&$ 100.0 $& IMBH &  1  & $40.88\pm 9.33$ &$0.05$ &$0.03$  \\
Djorg2	& $34.5 $&$ 0.0 $&$ 65.5 $& IMBH &  1  & $8.00\pm 1.94$ &$1.08$ &$0.54$  \\
NGC6517	& $61.8 $&$ 10.9 $&$ 27.3 $& NONE &  -  &  - &$3.41$ &$1.71$  \\
Terzan10	& $49.1 $&$ 0.0 $&$ 50.9 $& IMBH &  1  & $16.75\pm 3.87$ &$0.59$ &$0.30$  \\
NGC6522	& $54.5 $&$ 0.0 $&$ 45.5 $& NONE &  -  &  - &$1.96$ &$0.98$  \\
NGC6535	& $50.9 $&$ 0.0 $&$ 49.1 $& NONE &  -  &  - &$0.14$ &$0.07$  \\
NGC6528	& $72.7 $&$ 0.0 $&$ 27.3 $& NONE &  -  &  - &$0.73$ &$0.36$  \\
NGC6539	& $32.7 $&$ 29.1 $&$ 38.2 $& UNCLEAR &  -  &  - &$3.54$ &$1.77$  \\
NGC6544	& $21.8 $&$ 0.0 $&$ 78.2 $& IMBH &  1  & $3.98\pm 1.03$ &$1.02$ &$0.51$  \\
NGC6541	& $83.6 $&$ 0.0 $&$ 16.4 $& NONE &  -  &  - &$4.38$ &$2.19$  \\
2MS-GC01	& $0.0 $&$ 0.0 $&$ 100.0 $& IMBH &  1  & $11.56\pm 2.73$ &$0.48$ &$0.24$  \\
ESO-SC06	& $98.2 $&$ 0.0 $&$ 1.8 $& NONE &  -  &  - &$0.15$ &$0.08$  \\
NGC6553	& $40.0 $&$ 12.7 $&$ 47.3 $& UNCLEAR &  -  &  - &$2.19$ &$1.10$  \\
2MS-GC02	& $0.0 $&$ 0.0 $&$ 100.0 $& IMBH &  1  & $8.90\pm 2.14$ &$0.15$ &$0.08$  \\
NGC6558	& $16.4 $&$ 0.0 $&$ 83.6 $& IMBH &  1  & $31.96\pm 7.28$ &$0.64$ &$0.32$  \\
{\color{blue} {\bf IC1276}}	& $21.8 $&$ 0.0 $&$ 78.2 $& IMBH &  1  & $21.64\pm 4.96$ &$0.80$ &$0.40$  \\
Terzan12	& $0.0 $&$ 0.0 $&$ 100.0 $& IMBH &  1  & $18.99\pm 4.37$ &$0.08$ &$0.04$  \\
{\color{purple} {\bf NGC6569}}	& $50.9 $&$ 40.0 $&$ 9.1 $& NONE &  -  &  - &$3.51$ &$1.75$  \\
BH261	& $0.0 $&$ 0.0 $&$ 100.0 $& IMBH &  1  & $18.30\pm 4.22$ &$0.08$ &$0.04$  \\
{\color{red} {\bf NGC6584}}	& $30.9 $&$ 69.1 $&$ 0.0 $& BHS & $69 \pm 15$ & $0.83\pm 0.19$ &$2.04$ &$1.02$  \\
NGC6624	& $52.7 $&$ 0.0 $&$ 47.3 $& NONE &  -  &  - &$1.69$ &$0.85$  \\
NGC6626	& $74.5 $&$ 0.0 $&$ 25.5 $& NONE &  -  &  - &$3.14$ &$1.57$  \\
NGC6638	& $23.6 $&$ 0.0 $&$ 76.4 $& IMBH &  1  & $5.05\pm 1.27$ &$1.21$ &$0.60$  \\
NGC6637	& $56.4 $&$ 38.2 $&$ 5.5 $& NONE &  -  &  - &$1.95$ &$0.97$  \\
NGC6642	& $87.3 $&$ 0.0 $&$ 12.7 $& NONE &  -  &  - &$0.79$ &$0.39$  \\
NGC6652	& $85.5 $&$ 0.0 $&$ 14.5 $& NONE &  -  &  - &$0.79$ &$0.39$  \\
{\color{dgreen} {\bf NGC6656}}	& $25.5 $&$ 40.0 $&$ 34.5 $& UNCLEAR &  -  &  - &$4.30$ &$2.15$  \\
Pal8	& $61.8 $&$ 0.0 $&$ 38.2 $& NONE &  -  &  - &$0.27$ &$0.14$  \\
NGC6681	& $20.0 $&$ 0.0 $&$ 80.0 $& IMBH &  1  & $7.17\pm 1.75$ &$1.21$ &$0.60$  \\
GLIMPSE01	& $34.5 $&$ 0.0 $&$ 65.5 $& IMBH &  1  & $5.01\pm 1.27$ &$0.40$ &$0.20$  \\
{\color{red} {\bf NGC6712}}	& $14.5 $&$ 58.2 $&$ 27.3 $& BHS & $70 \pm 15$ & $0.84\pm 0.19$ &$1.71$ &$0.86$  \\
NGC6715	& $98.2 $&$ 1.8 $&$ 0.0 $& NONE &  -  &  - &$16.79$ &$8.39$  \\
NGC6717	& $45.5 $&$ 0.0 $&$ 54.5 $& IMBH &  1  & $11.56\pm 2.73$ &$0.31$ &$0.16$  \\
{\color{dgreen} {\bf NGC6723}}	& $45.5 $&$ 21.8 $&$ 32.7 $& UNCLEAR &  -  &  - &$2.32$ &$1.16$  \\
NGC6749	& $52.7 $&$ 0.0 $&$ 47.3 $& NONE &  -  &  - &$0.82$ &$0.41$  \\
NGC6752	& $61.8 $&$ 0.0 $&$ 38.2 $& NONE &  -  &  - &$2.11$ &$1.06$  \\
NGC6760	& $41.8 $&$ 45.5 $&$ 12.7 $& UNCLEAR &  -  &  - &$2.34$ &$1.17$  \\
{\color{red} {\bf NGC6779}}	& $30.9 $&$ 65.5 $&$ 3.6 $& BHS & $81 \pm 17$ & $0.98\pm 0.22$ &$1.57$ &$0.79$  \\
Terzan7	& $60.0 $&$ 0.0 $&$ 40.0 $& NONE &  -  &  - &$0.17$ &$0.09$  \\
Pal10	& $0.0 $&$ 0.0 $&$ 100.0 $& IMBH &  1  & $13.56\pm 3.17$ &$0.35$ &$0.18$  \\
Arp2	& $92.7 $&$ 7.3 $&$ 0.0 $& NONE &  -  &  - &$0.22$ &$0.11$  \\
{\color{red} {\bf NGC6809}}	& $5.5 $&$ 94.5 $&$ 0.0 $& BHS & $108 \pm 23$ & $1.34\pm 0.30$ &$1.82$ &$0.91$  \\
Terzan8	& $69.1 $&$ 21.8 $&$ 9.1 $& NONE &  -  &  - &$0.18$ &$0.09$  \\
{\color{dgreen} {\bf Pal11}}	& $47.3 $&$ 49.1 $&$ 3.6 $& UNCLEAR &  -  &  - &$1.00$ &$0.50$  \\
NGC6838	& $0.0 $&$ 0.0 $&$ 100.0 $& IMBH &  1  & $17.67\pm 4.08$ &$0.30$ &$0.15$  \\
NGC6864	& $100.0 $&$ 0.0 $&$ 0.0 $& NONE &  -  &  - &$4.58$ &$2.29$  \\
{\color{dgreen} {\bf NGC6934}	}& $41.8 $&$ 47.3 $&$ 10.9 $& UNCLEAR &  -  &  - &$1.63$ &$0.82$  \\
{\color{dgreen} {\bf NGC6981}}	& $36.4 $&$ 29.1 $&$ 34.5 $& UNCLEAR &  -  &  - &$1.12$ &$0.56$  \\
NGC7006	& $21.8 $&$ 60.0 $&$ 18.2 $& BHS & $120 \pm 25$ & $1.52\pm 0.34$ &$2.00$ &$1.00$  \\
NGC7078	& $89.1 $&$ 0.0 $&$ 10.9 $& NONE &  -  &  - &$8.11$ &$4.06$  \\
NGC7089	& $76.4 $&$ 9.1 $&$ 14.5 $& NONE &  -  &  - &$7.00$ &$3.50$  \\
NGC7099	& $72.7 $&$ 0.0 $&$ 27.3 $& NONE &  -  &  - &$1.63$ &$0.82$  \\
Pal12	& $0.0 $&$ 0.0 $&$ 100.0 $& IMBH &  1  & $232.65\pm 59.21$ &$0.10$ &$0.05$  \\
Pal13	& $16.4 $&$ 0.0 $&$ 83.6 $& IMBH &  1  & $75.41\pm 17.57$ &$0.05$ &$0.03$  \\
NGC7492	& $70.9 $&$ 16.4 $&$ 12.7 $& NONE &  -  &  - &$0.36$ &$0.18$  \\
\hline \hline
\hline
\end{longtable}

\bsp	
\label{lastpage}
\end{document}